\documentclass[twocolumn,twoside,a4paper]{revtex4}

\usepackage[ansinew]{inputenc}
\usepackage[T1]{fontenc}
\usepackage[english]{babel}
\usepackage[dvips]{graphicx}
\usepackage{amsmath}
\usepackage{amssymb}
\usepackage{amsfonts}

\renewcommand{\vec}{\mathbf}

\begin{document}

\title{Venus-Solar Wind Interaction: Asymmetries and the Escape of O$^+$ Ions}
\author{E. Kallio}
\email{esa.kallio@fmi.fi}
\author{R. Jarvinen}
\email{riku.jarvinen@fmi.fi}
\affiliation{Finnish Meteorological Institute, Space Research Unit}
\author{P. Janhunen}
\email{pekka.janhunen@fmi.fi}
\affiliation{University of Helsinki, Department of Physical Sciences}
\altaffiliation{Also at Finnish Meteorological Institute}

\begin{abstract}
We study the interaction between Venus and the solar wind using a global three-dimensional self-consistent quasi-neutral hybrid (QNH) model. The model treats ions (H$^+$, O$^+$) as particles and electrons as a massless charge neutralising fluid. In the analysed Parker spiral interplanetary magnetic field (IMF) case (IMF = [8.09, 5.88, 0] nT) a notable north-south asymmetry of the magnetic field and plasma exists, especially in the properties of escaping planetary O$^+$ ions. The asymmetry is associated with ion finite gyroradius effects. Furthermore, the IMF $x$-component results in a dawn-dusk asymmetry. Overall, the QNH model is found to reproduce the main observed plasma and magnetic field regions (the bow shock, the magnetosheath, the magnetic barrier and the magnetotail), implying the potential of the developed model to study the Venusian plasma environment and especially the non-thermal ion escape.
\end{abstract}

\maketitle

\section{Introduction}
Neither Venus nor Mars has a strong global intrinsic magnetic field and therefore the solar wind is able to flow close to the planets in regions where the exospheric neutral density is high. Because of the resulting direct interaction between the exosphere and the solar wind, ionized atmospheric neutrals are accelerated (picked up) by the solar wind electric field. Charge exchange reactions between the solar wind protons and planetary neutrals also produce energetic neutral hydrogen atoms (H-ENAs). Observation of H-ENAs is thus a manifestation of the direct interaction between the solar wind plasma and planetary neutrals. Picked-up planetary O$^+$ ions can also form energetic neutral oxygen atoms (O-ENA) via charge exchange process. The ion escape, H-ENAs, O-ENAs and electrons will be investigated at Venus and Mars by two identical instruments: ASPERA-4 on Venus Express \citep[see][]{BarabashEtAl_ASPERA4_2006} and ASPERA-3 on Mars Express \citep[see][]{BarabashEtAl_ASPERA3_2004}.

The Venus-solar wind interaction can be analysed self-consistently by three-dimensional (3-D) magnetohydrodynamic (MHD) and quasi-neutral hybrid (QNH) models. The MHD models \citep[see, for example,][and references therein]{KallioEtAl_VenusMHD_1998} are computationally cheaper than the QNH models \citep[see, for example,][]{BrechtFerrante_Hybrid_1991,Shimazu_Hybrid_1999}, thus allowing a better grid resolution. Two-dimensional (2-D) QNH models \citep{TeradaEtAl_Hybrid_2004} allow similar or even better resolution than 3-D MHD models, but then e.g. the 3-D draping of the magnetic field around the planet cannot be efficiently studied. An advantage of a 3-D multi-ion species QNH model, as compared with the 3-D multifluid MHD model, is that it includes finite ion gyroradius effects. Furthermore, different ion species are allowed to have different velocities and temperatures. Even arbitrary non-Maxwellian distributions functions are allowed. Therefore, the QNH model makes it possible to increase our understanding of those plasma environments where the ion gyroradius is comparable to or larger than the planet size.

In this paper the Venusian plasma environment is studied with a recently developed 3-D QNH model. The model is based on previous 3-D QNH models that have been used to study plasma and magnetic fields near Mars \citep{Kallio_Janhunen_Mars_2002}, Mercury \citep{KallioJanhunen_Mercury_2003}, Saturnian moon Titan \citep{KallioEtAl_Titan_2004} and the Moon \citep{Kallio_Moon_2005}. The analysis focuses on the asymmetries caused by kinetic effects, the role of the IMF (interplanetary magnetic field) $x$-component and the properties of the escaping planetary O$^+$ ions.

The paper is organized as follows. First the developed Venus model is described. Then the basic properties of H$^+$ ions, O$^+$ ions and the magnetic field near Venus are presented.

\section{Model Description}
\subsection{Quasi-neutral hybrid model (QNH)}
The quasi-neutral hybrid model (QNH) is based on the QNH models that have been used earlier to study Mars, Mercury, Titan, and the Moon \citep{Kallio_Janhunen_Mars_2002,KallioJanhunen_Mercury_2003,KallioEtAl_Titan_2004,Kallio_Moon_2005}. Here we describe only some of its basic features and its new possibilities.

In this work, the coordinate system is a Cartesian system with the origin fixed at the centre of Venus. The $x$-axis points from Venus towards the Sun. The $z$-axis is defined by $\vec{u}_{z}=\vec{u}_{x}\times\vec{v}/ |\vec{u}_{x}\times\vec{v}|$, where $\vec{u}_{x}$ is a unit vector in the positive $x$-direction, and $\vec{v}$ is the orbital velocity of Venus around the Sun, i.e., $z$ is perpendicular to the orbital plane of Venus. The $y$-axis completes the right-handed coordinate system.

\begin{figure}
 \begin{center}
  \includegraphics[width=0.49\textwidth]{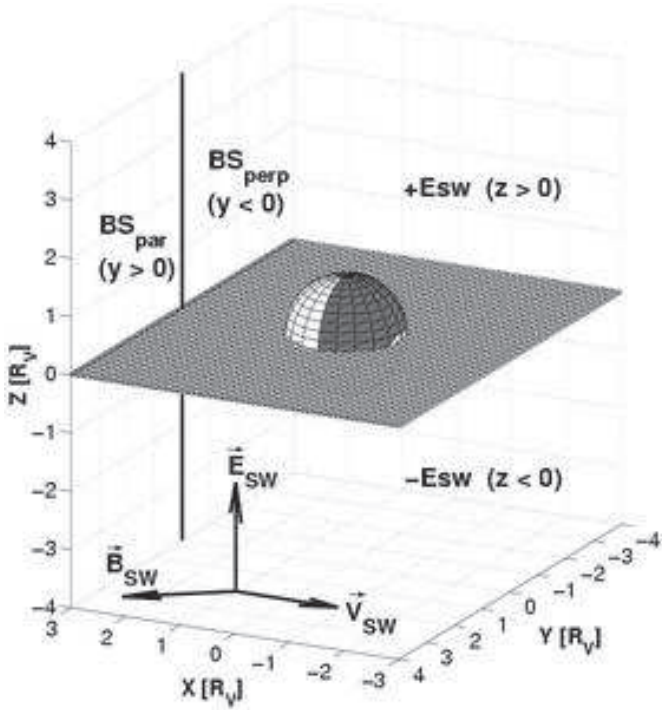}
  \caption{Illustration of the used simulation box and the adopted terminology. See text for details.} \label{pic:grid}
 \end{center}
\end{figure}

Figure \ref{pic:grid} illustrates the coordinate system, grid structure and terminology employed. The size of the simulation box is $-3 R_V < x < 3 R_V$ and $-4 R_V < y,z < 4 R_V$ and the radius of the obstacle to the plasma flow is taken to be $R_V = 6051.8$ km. The grid size is one tenth of the obstacle radius (605 km). The upstream plasma parameters analysed in this paper are $n_\textrm{SW}$ = 14 cm$^{-3}$, $\vec{U}_\textrm{SW}$ = [-430, 0, 0] km s$^{-1}$ and the thermal velocity 50 km s$^{-1}$. The IMF corresponds to a Parker spiral angle (= $\arctan(\sqrt{B_y^2 + B_z^2}/|B_x|)$ of 36$^\circ$: $\vec{B}_\textrm{SW}$ = [cos(36$^\circ$), sin(36$^\circ$), 0] 10 nT = [8.09, 5.88, 0] nT.

It should be noted that the aberration caused by the orbital motion of Venus of about 35 km s$^{-1}$ caused by the rotation of the planet around the Sun had not been taken into account. In the used solar wind velocity of 430 km s$^{-1}$ the aberration angle would have been about 5$^\circ$ ($=\arctan$(35 km s$^{-1}$ / 430 km s$^{-1}$)). Had the aberration been taken into account, the solution presented in this paper would have been rotated about 5$^\circ$ clockwise around the +$z$-axis if the results were presented in a coordinate system where $x$-axis points from Venus towards the Sun. In the rotation, the IMF $x$ and $y$ components would also have been slightly changed.

As depicted in Fig. \ref{pic:grid}, in this paper the $z > 0$ hemisphere is referred to as the +Esw hemisphere. The convective electric field in the solar wind $\vec{E}_\textrm{SW}$ ($=-\vec{U}_\textrm{SW} \times \vec{B}_\textrm{SW}$) on the $XY$ plane points toward the +Esw hemisphere.  The opposite hemisphere is referred to as the -Esw hemisphere. The terminology reminds us of the fact that the results presented depend on the the choice of the direction of the IMF and that the asymmetries rotate around the $x$-axis along with the IMF.  The +Esw and -Esw hemispheres are important when the properties of the O$^+$ ions are considered, as will be seen in the next section.

In addition, the $y < 0$ hemisphere is called the BS$_{perp}$ hemisphere because there are points on the bow shock on the $y < 0$ hemisphere where the IMF is perpendicular and quasi-perpendicular to the bow shock in the analysed IMF case. Similarly, the opposite $y > 0$ hemisphere is referred to as the BS$_{par}$ because there are points on the bow shock on the $y > 0$ hemisphere where the IMF is parallel and quasi-parallel to the bow shock in the analysed IMF case. It should be noted that if one had analysed the IMF $B_x < 0$ case the BS$_{perp}$ hemisphere would have been at the $y > 0$ hemisphere. The BS$_{perp}$/BS$_{par}$ hemispheres will be discussed in the next section when the asymmetries in the magnetosheath and the magnetic tail lobes are considered.

The model contains two ion species, H$^+$ and O$^+$. The ions are modelled as particles and they are accelerated by the Lorentz force:
\begin{eqnarray}
 m_{\textrm{H}^+} \frac{d \vec{v}_{\textrm{H}^+}}{dt} & = & e ( \vec{E} + \vec{v}_{\textrm{H}^+} \times \vec{B})\label{eq:lorentz_H}\\
 m_{\textrm{O}^+} \frac{d \vec{v}_{\textrm{O}^+}}{dt} & = & e ( \vec{E} + \vec{v}_{\textrm{O}^+} \times \vec{B}).\label{eq:lorentz_O}
\end{eqnarray}
Here $\vec{v}$, $m$, $\vec{E}$, $\vec{B}$ and $e$ are the ion velocity, ion mass, electric field, magnetic field and elementary charge, respectively. The subscript H$^+$(O$^+$) refers to protons(oxygen ions).

Electrons form a massless fluid and their equation of motion (the momentum equation) is
\begin{eqnarray} \label{eq:ele}
 \vec{E} + \vec{U}_\textrm{e} \times \vec{B} & = & \eta_a \vec{J},
\end{eqnarray}
where $n_\textrm{e}$, $\vec{U}_\textrm{e}$, $\vec{J}$ and $\eta_a$ are the electron density, electron bulk velocity, electric current density and the anomalous resistivity, respectively. In the hybrid model the resistivity function can be a fully 3-D function, i.e.,$\eta_a = \eta_a(x,y,z)$. At Venus the resistivity is largest near the planet, where the electron-neutral collision frequency becomes high, and smallest in the undisturbed solar wind. In this paper, however, no $\it{ad~hoc}$ 3-D or 2-D resistivity models are developed. In contrast, a constant resistivity 230 k$\Omega$m is used everywhere outside Venus for simplicity.

The electric current is defined as
\begin{eqnarray} \label{eq:J}
 \vec{J} & = & e \big( n_{\textrm{H}^+} \vec{V}_{\textrm{H}^+} + n_{\textrm{O}^+} \vec{V}_{\textrm{O}^+} - n_\textrm{e} \vec{U}_\textrm{e} \big).
\end{eqnarray}
Here $n_i$ and $\vec{V}_i$ are the number density and average velocity of the $i$th ion species. The electric current density $\vec{J}$ is related to the magnetic field $\vec{B}$ according to Amp\`{e}re's law:
\begin{eqnarray}
 \nabla \times \vec{B} =  \mu_0 \vec{J}\label{eq:ampere},
\end{eqnarray}
where $\mu_0$ is the vacuum permeability.

The QNH model assumes quasi-neutrality, i.e.,
\begin{eqnarray} \label{eq:n_e}
 n_\textrm{e}  =  n_{\textrm{H}^+} + n_{\textrm{O}^+}.
\end{eqnarray}

The magnetic field is propagated by Faraday's law
\begin{eqnarray}
 \frac{\partial \vec{B}}{\partial t} & = & - \nabla \times \vec{E}, \label{eq:faraday}
\end{eqnarray}
where the electric field is derived from Eq.~$\ref{eq:ele}$.

The solution of Equations \ref{eq:lorentz_H} - \ref{eq:faraday} proceeds as follows. First, the total current density is derived from the magnetic field using Amp\`{e}re's law (\ref{eq:ampere}). Second, the velocity field of the electron fluid is derived from the electric current, the number density and the average speed of ions by Equations \ref{eq:J} and \ref{eq:n_e}. Third, the updated electric field is derived from the updated magnetic field, the current density and the velocity field of the electron fluid by using Equation \ref{eq:ele}. Finally, the ions are moved according to Equations \ref{eq:lorentz_H} and \ref{eq:lorentz_O} by using the new electric and magnetic fields. The movement of ions results in updated plasma densities and bulk velocities and the magnetic field can be propagated by using the updated electric field in Faraday's law (\ref{eq:faraday}) \citep[see][for the details of the used algorithms]{KallioJanhunen_Mercury_2003}.

The obstacle within the simulation box is assumed to represent an ideally conducting ball inside which the resistivity is zero. We apply fully absorbing boundary conditions on the surface of the obstacle by taking an ion away from the simulation if it crosses the obstacle boundary. In the analyzed run the simulation box contains about 12 million H$^+$ ions and about 0.2 million O$^+$ ions at every time step. The average number of macroions per a cell is 30 and the timestep $\Delta t = $ 0.04 s.

In the model the interplanetary magnetic field is derived from the velocity field that represents the flow of a laminar incompressible fluid around a spherical obstacle by assuming that the magnetic field is "frozen in" to the laminar incompressible flow. The same approach has been used earlier in a 3-D QNH model \citep{Shimazu_Hybrid_2001a}. If the flow at infinity is along the (negative) $x$-axis, the magnetic field outside of the obstacle can be shown to be
\begin{eqnarray}
 \widetilde{B}_x & = & B_0 \bigg( 1 + \frac{R^3_\textrm{V}}{2 r^3} -\frac{3 R^3_\textrm{V} x^2}{2 r^5} \bigg)\nonumber\\
 \widetilde{B}_y & = & -  B_0 \frac{3 R^3_\textrm{V} x y}{2 r^5}\label{eq:laminarB}\\
 \widetilde{B}_z & = & - B_0 \frac{3 R^3_\textrm{V} x z}{2 r^5}\nonumber,
\end{eqnarray}
where $B_0$ is the magnitude of the IMF and $r = \sqrt{x^2 + y^2 + z^2}$. The radial component of the magnetic field on the surface of the obstacle is zero. The total magnetic field in the simulation box outside of the obstacle is therefore
\begin{eqnarray} \label{eq:Btotal}
 \vec{B}(\vec{r},t)  = \widetilde{\vec{B}}_{laminar}(\vec{r}) + \delta \vec{B}(\vec{r},t),
\end{eqnarray}
where $\widetilde{\vec{B}}_{laminar}(\vec{r})$ is time independent magnetic field derived from Eq. \ref{eq:laminarB} and $\delta \vec{B}(\vec{r},t)$ is time dependent self-consistent magnetic field. It should be noted that $\widetilde{\vec{B}}_{laminar}(\vec{r})$ is derived from the potential field ($\widetilde{\vec{B}}_{laminar} = \nabla(f)$) and therefore there are no electric currents associated with the $\widetilde{\vec{B}}_{laminar}(\vec{r})$ field outside the obstacle ($\vec{J} \sim \nabla \times \nabla(f)$ = 0). The magnetic field component $\widetilde{\vec{B}}_{laminar}(\vec{r})$ is associated with electric currents on the surface of the obstacle.

Equations \ref{eq:laminarB} are needed to introduce the IMF $x$-component into the simulation box. In the previous 3-D QNH Venus model the IMF $x$-component has been assumed to be zero \citep{BrechtFerrante_Hybrid_1991,Shimazu_Hybrid_2001a} which is not a good approximation at Venus where the nominal Parker spiral angle is about 36$^\circ$ and, consequently, the $x$-component is often larger in magnitude that the $y$ or $z$ components. Implementation of the IMF $x$-component enables us to study its associated set of asymmetries.

\begin{figure}
 \begin{center}
  \includegraphics[width=0.49\textwidth]{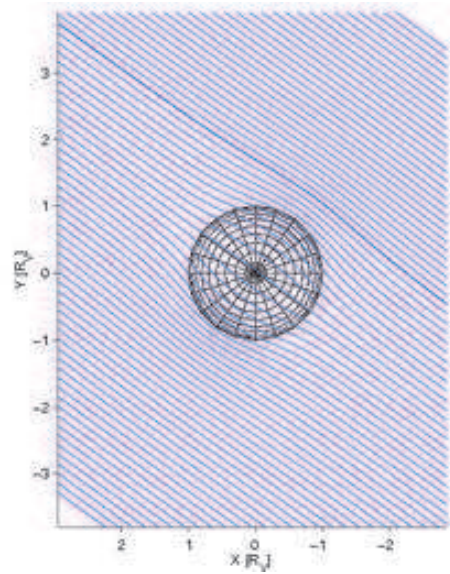}
  \caption{An example of magnetic field which is "frozen in" to the laminar incompressible flow around a sphere. Such a constant magnetic field is used in this work to generate the IMF $x$- and $y$-components.} \label{pic:IMF_potential}
 \end{center}
\end{figure}

In this work Equations \ref{eq:laminarB} are used to model the full IMF, and not only its $x$-component, by rotating the laminar flow around the $z$-axis in order to produce a magnetic field which is equal to the IMF at infinity. Such a magnetic field is derived from Equations \ref{eq:laminarB} by rotating the magnetic field around the $z$-axis. Figure \ref{pic:IMF_potential} shows an example of the magnetic fields used in this work.

\subsection{Plasma sources}
The simulations contain three H$^+$ sources: the solar wind, the cold atmospheric hydrogen and the hot neutral hydrogen exosphere. Likewise there are two O$^+$ sources: the cold atmospheric oxygen and the hot neutral oxygen exosphere. The major source for H$^+$ is the solar wind. In the simulation the solar wind H$^+$ ions are injected to the simulation box at the front face at $x = 3 R_V$. The planetary hydrogen and oxygen densities are similar, or slightly simplified versions, of the density profiles that have been used recently to study generation of ENAs at Venus \citep{GunellEtAl_ENA_2005}.

\subsubsection{Hydrogen atoms}
The cold atmospheric hydrogen is modelled by the Chamberlain exosphere model
\begin{eqnarray} \label{eq:Chamberlain}
 n(r) & = & n_0 e^{-\beta (1/r_0 - 1/r)} \zeta(\frac{\beta}{r}),
\end{eqnarray}
where $n_0$ is the hydrogen density at the planetocentric distance $r = r_0$, $\zeta$ is Chamberlain's partition function and
\begin{eqnarray}
 \beta & = & \frac{G M_\textrm{V} m_{\textrm{H}^+}}{k_B T}.
\end{eqnarray}
Here $G$, $M_\textrm{V}$, $k_B$ and $T$ are the gravitational constant, planet's mass, Boltzmann's constant and neutral temperature, respectively.  At noon we adopt the parameters $n_0 = 1.32 \times 10^{11} \textrm{ m}^{-3}$ and $T$ = 285 K and at the midnight $n_0 = 2.59 \times 10^{12} \textrm{ m}^{-3}$ and $T = 110$ K. Both noon and midnight have $r_0 = 170 \textrm{ km} + R_V$. The densities between noon and midnight are obtained by linear interpolation with respect to the solar zenith angle (SZA). Chamberlain's partition function differs only slightly from unity within the simulation box and, therefore, it's value was taken to be unity.

The hot hydrogen exosphere (corona) is modelled at noon, the terminator and midnight by a function
\begin{eqnarray} \label{eq:H_hot_model}
 n(r) & = & e^{a_1 r + a_2 + a_3/r},
\end{eqnarray}
where the coefficients $a_i$ are given in Table \ref{tbl:H_hot_coef}. Linear interpolation in SZA is used to get densities between SZA = 0$^\circ$, 90$^\circ$ and 180$^\circ$.
\begin{table}[h]
 \begin{center}
  \begin{tabular}{|c|c|c|c|}
   \hline
   SZA (deg.) & $a_1$ [km$^{-1}$] & $a_2$ & $a_3$ [km]\\
   \hline
   0 & $-6.2625 \times 10^{-5}$ & 15.4817 & $3.6414 \times 10^4$\\
   $90$ & $-8.4607 \times 10^{-5}$ & 15.9944 & $2.9743 \times 10^4$\\
   $180$ & $-6.2309 \times 10^{-5}$ & 15.2723 & $4.3781 \times 10^4$\\
   \hline
  \end{tabular}
  \caption{The constants used in Equation \ref{eq:H_hot_model} to model the hot hydrogen exosphere (corona).}\label{tbl:H_hot_coef}
 \end{center}
\end{table}

\subsubsection{Oxygen atoms}
The atmospheric scale height of the cold oxygen population is much smaller than the grid size on the simulation (= 605 km) and therefore the emission of oxygen ions originating from the cold oxygen component is modelled by emitting oxygen ions from the obstacle boundary. The particle flux has its maximum at the noon and it decreases as $\cos(\textrm{SZA})$ towards the terminator. At the nightside the oxygen ion emission rate is 10\% of the noon value. The total O$^+$ emission rate from the obstacle boundary is $10^{25} \textrm{ s}^{-1}$.

The hot oxygen exosphere is modelled by Equation \ref{eq:Chamberlain} with the following values: at noon $n_0 = 7.5 \times 10^{10} \textrm{ m}^{-3}$, $r_0 = 200$ km and $T = 6400$ K and at midnight $n_0 = 2 \times 10^9 \textrm{ m}^{-3}$, $r_0 = 300$ km and $T = 4847$ K. Values for other SZA values are obtained by linear interpolation.

\subsubsection{Ionization}
The  hydrogen ion production rate $q_{\textrm{H}^+}$ [$\textrm{s}^{-1}$ $\textrm{m}^{-3}$] and the oxygen ion production rate $q_{\textrm{O}^+}$ [$\textrm{s}^{-1}$ $\textrm{m}^{-3}$] in sunlight are modelled as
\begin{eqnarray} \label{eq:Ionization}
 q_{\textrm{H}^+}(\vec{r}) & = & f^{tot}_{\textrm{H}^+}(\vec{r}) ~ n_\textrm{H}(\vec{r}),\\
 q_{\textrm{O}^+}(\vec{r}) & = & f^{tot}_{\textrm{O}^+}(\vec{r}) ~ n_\textrm{O}(\vec{r}),
\end{eqnarray}
where $f^{tot}_{\textrm{H}^+}(\vec{r})$($f^{tot}_{\textrm{O}^+}(\vec{r})$) [$\textrm{s}^{-1}$] is the total ionization frequency and $n_\textrm{H}(\vec{r})$ ($n_\textrm{O}(\vec{r})$) [m$^{-3}$] the total density of the hydrogen(oxygen) atoms at $\vec{r}$. $f^{tot}_{\textrm{H}^+}$ and $f^{tot}_{\textrm{O}^+}$ are zero within the optical shadow. The three most important ionization processes at Venus are photoionisation, electron impact ionisation and charge exchange \citep[see, for example,][]{ZhangEtAl_CX_1993}. In this work only EUV ionization is taken into account for simplicity. Table \ref{tbl:ioni_coeff} shows the used constant photoionisation frequencies for solar minimum conditions and the corresponding total ion EUV production rates within the simulation box.

\begin{table}[h]
 \begin{center}
  \begin{tabular}{|c|c|c|c|r|}
   \hline
   Source & Ionization freq. & Total rate & Comments \\
          & $f$ (1/s)        & (\#/s)     & \\
   \hline
   H$_{ \genfrac{}{}{0pt}{}{ \textrm{hot} }{ \textrm{cold} } }$ $\rightarrow$ H$^+$ &  $1.39 \times 10^{-7}$ & $6.42 \times 10^{24}$ & EUV ionization\\
   O$_\textrm{hot}$ $\rightarrow$ O$^+$ & $4.55 \times 10^{-7}$ & $4.09 \times 10^{24}$ & EUV ionization\\
   O$_\textrm{cold}$ $\rightarrow$ O$^+$ & - & $10^{25}$ & From the obstacle\\
   \hline
  \end{tabular}
  \caption{Summary of H$^+$ and O$^+$ production rates from the hot H neutral corona and cold hydrogen component and from the hot O neutral corona. The photoionization frequencies ($f$) are for the quiet solar EUV condition. Oxygen ions from the cold O$^+$ component is modelled by emitting oxygen ions from the obstacle boundary. Note that part of the generated oxygen ions hit to the obstacle boundary and that the total escape rate from the simulation box is about $2.2 \times 10^{24} \textrm{ s}^{-1}$.}\label{tbl:ioni_coeff}
 \end{center}
\end{table}

\begin{figure}
 \begin{center}
  \includegraphics[width=0.49\textwidth]{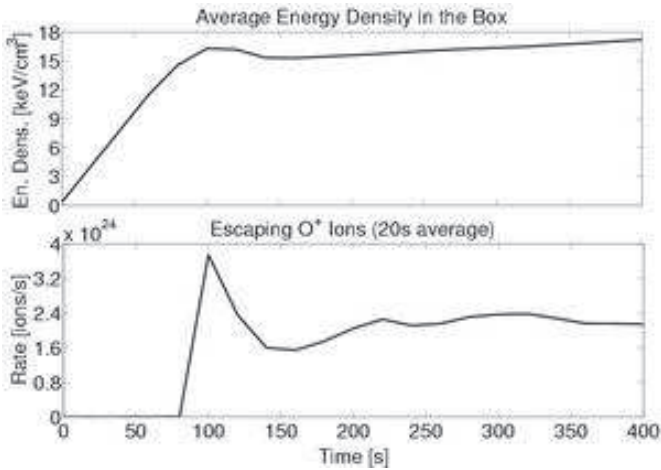}
  \caption{(a) The time evolution of the total energy density (kinetic energy density + magnetic energy density) inside the simulation box. (b) The number of escaping O$^+$ ions from the side faces of the simulation box.} \label{pic:plot1}
 \end{center}
\end{figure}

The presented run corresponds to the situation after 240 seconds from the start of the simulation. At that time the simulation has reached a quasi-stationary state where the total energy within the simulation box does not increase much (Figure \ref{pic:plot1}a). Also, at that time the total escape rate of O$^+$ ions from the simulation box has reached a relatively stationary rate of about $2.2 \times 10^{24} \textrm{ s}^{-1}$ (Fig. \ref{pic:plot1}b).

\section{Results}
In this sections several plasma and magnetic field parameters are presented on the $XY$ and $XZ$ planes. The used grid size of 0.1 $R_V$ prevents us from a detailed study of the position and the shape of the various boundary layers. Nevertheless, a conic shape is included in the figures to help the eye in catching asymmetries more easily and to simplify comparisons between various plasma and magnetic field parameters. The conical shape \citep{SlavinHolzer_BowShock_1981}

\begin{eqnarray}
 r & = & \frac{L}{1 + \epsilon \cos(\Theta)}, \label{eq:bs_r}\\
 \vec{r}_{focus} & = & (x_0,0,0)\label{eq:bs_x_0}
\end{eqnarray}
is used to provide a rough estimation for the shape of the bow shock. Here $r$ is the radial distance from the focus point, $\vec{r}_{focus}$ to the bow shock, $\epsilon$ is the eccentricity, $\Theta$ is the angle between the a point of a bow shock and the $x$-axis and $L$ is the distance of the bow shock from the focus point on the $x = x_0$ plane. In this paper the conical shape on the $XY$ and $XZ$ planes is derived from Equations \ref{eq:bs_r} - \ref{eq:bs_x_0} for the parameters $x_0 = 0.6 R_V$, $\epsilon = 0.87$ and $L = 1.8 R_V$.

\begin{figure*}
 \begin{center}
  \includegraphics[width=0.24\textwidth]{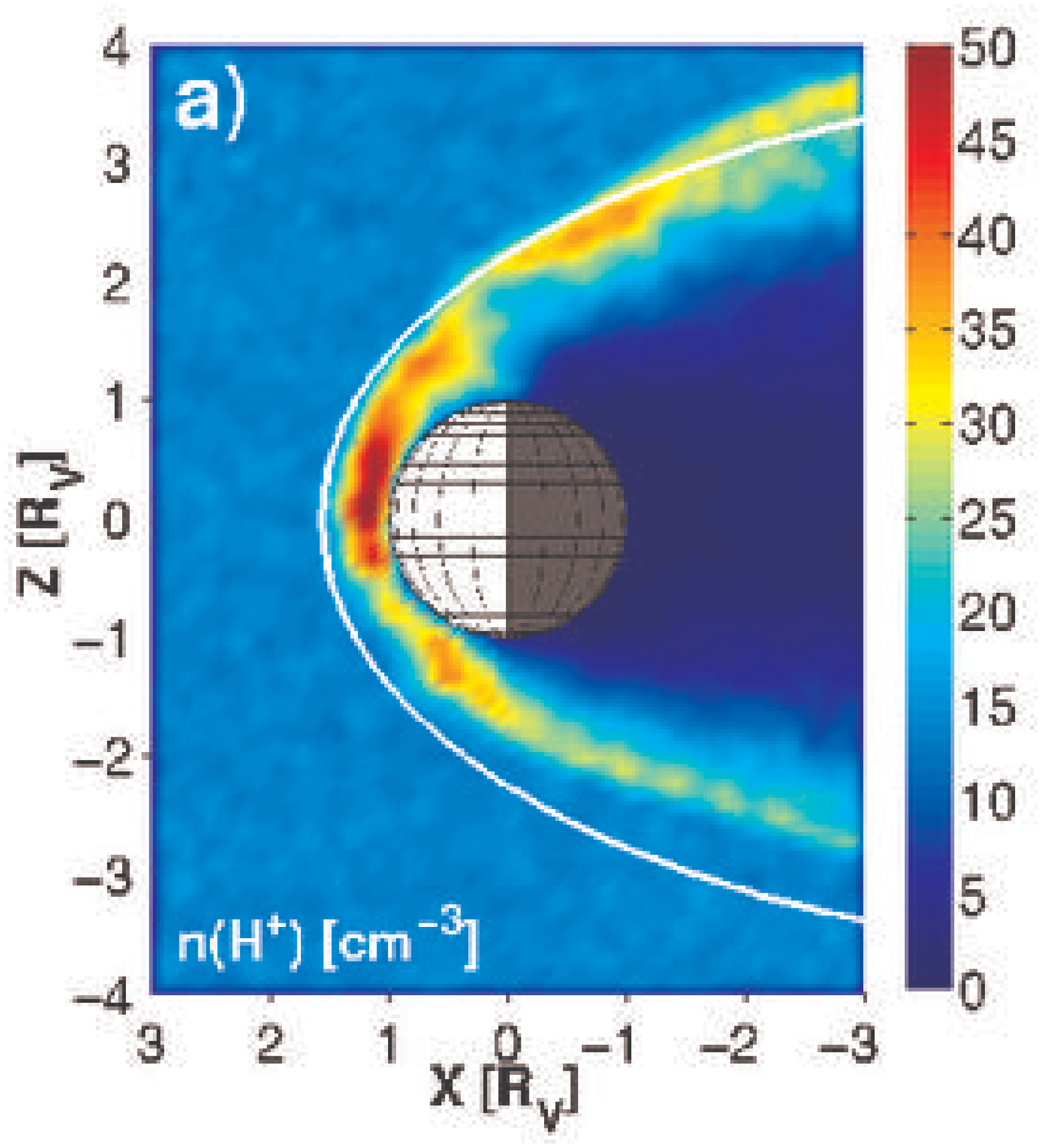}
  \includegraphics[width=0.24\textwidth]{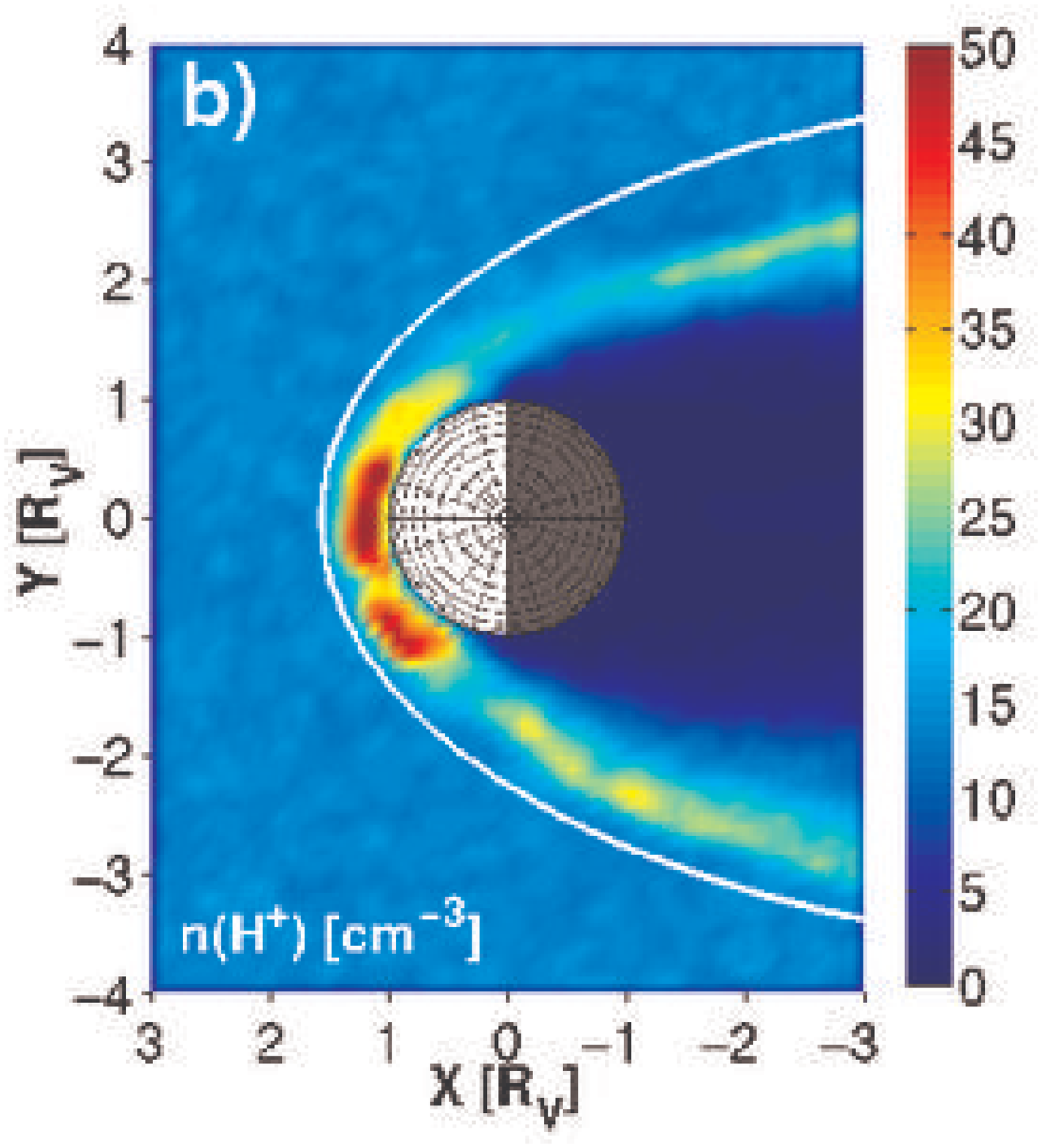}\\
  \includegraphics[width=0.24\textwidth]{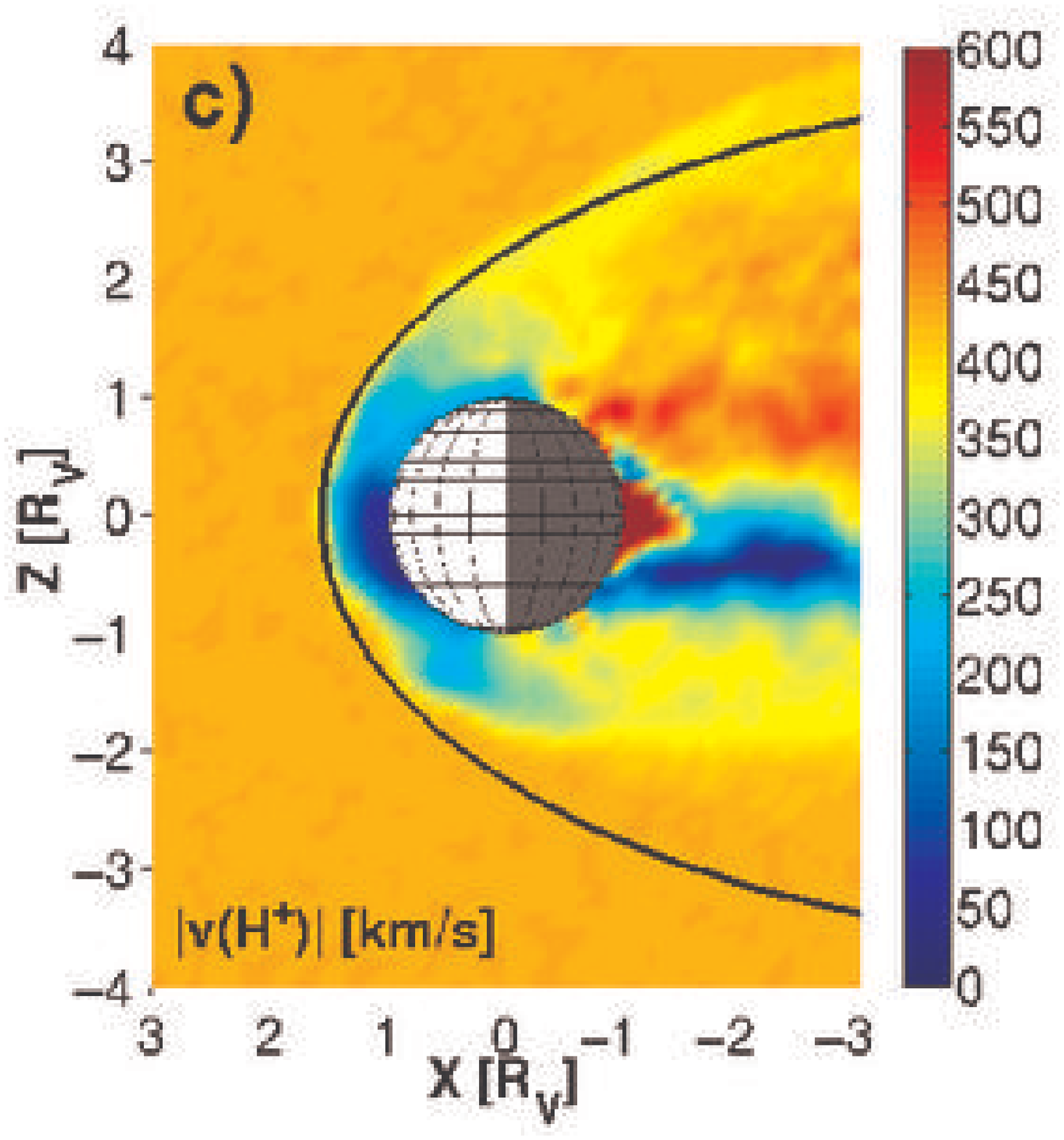}
  \includegraphics[width=0.24\textwidth]{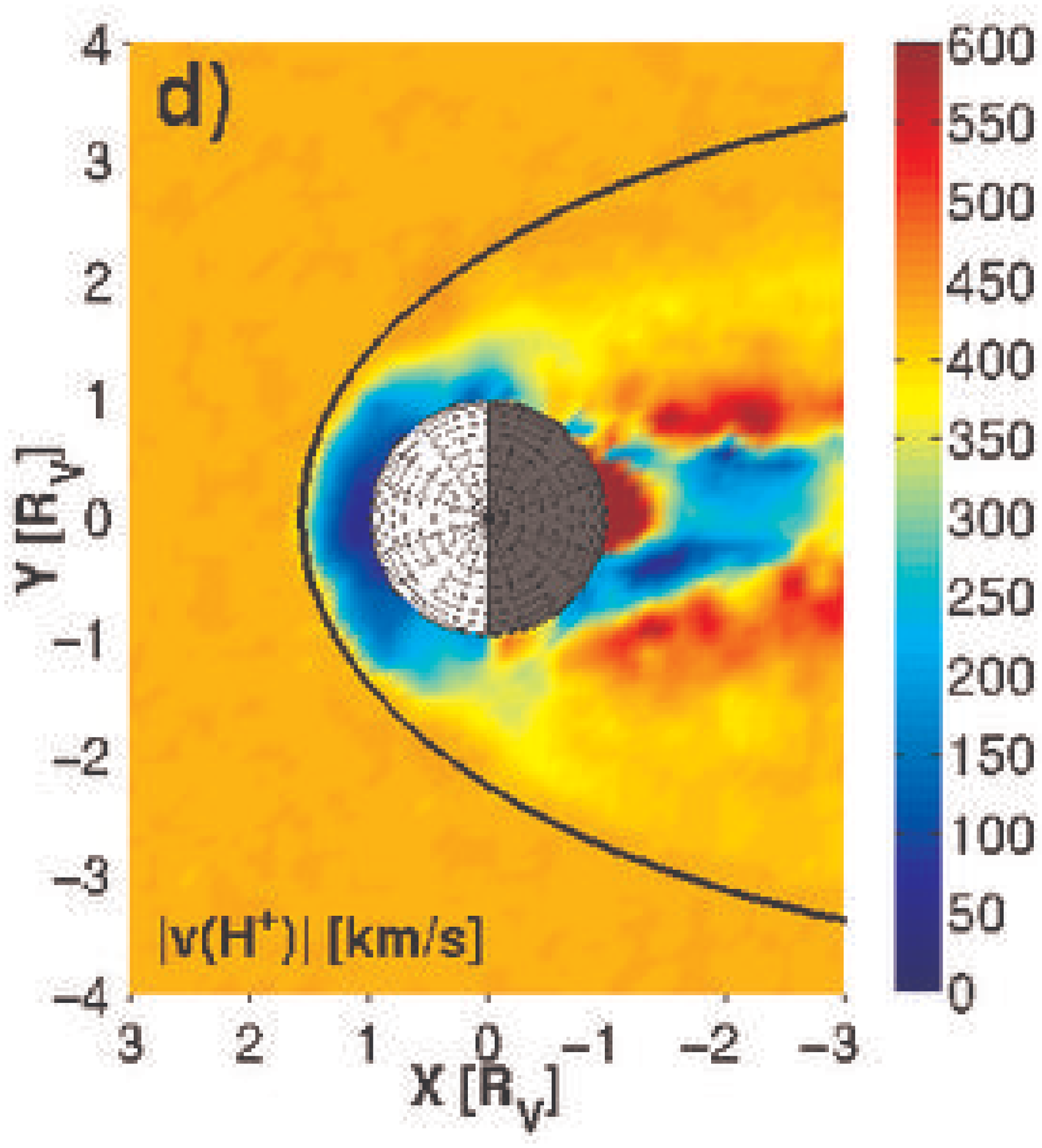}
  \caption{Properties of the H$^+$ ions near Venus: proton density [cm$^{-3}$] on (a) the $XZ$ plane and on (b) the $XY$ plane and the bulk velocity [km s$^{-1}$] on (c) the $XZ$ plane and on (d) the $XY$ plane.} \label{pic:H_xz_xy}
 \end{center}
\end{figure*}

Figure \ref{pic:H_xz_xy} presents the density and bulk velocity of protons in the $XZ$ and $XY$ planes. The proton density has increased and the velocity decreased at the bow shock. At the nightside behind the planet the density decreases again, becoming smaller than the undisturbed solar wind density. A clear asymmetry can be identified between the +Esw hemisphere ($z > 0$) and the -Esw hemisphere ($z < 0$).

It can be seen in Figs. \ref{pic:H_xz_xy}a and \ref{pic:H_xz_xy}c that the bow shock is more blunt and further away from the planet on the +Esw hemisphere than on the -Esw hemisphere. Furthermore, the bulk velocity on the nightside in the tail is higher on the +Esw hemisphere than on the opposite hemisphere. Only a slight asymmetry can be found between the BS$_{perp}$ hemisphere ($y < 0$) and on the opposite BS$_{par}$ hemisphere ($y > 0$). Note also that the distance of the bow shock from the planet is not identical in the $XY$ and $XZ$ planes.

\begin{figure*}
 \begin{center}
  \includegraphics[width=0.24\textwidth]{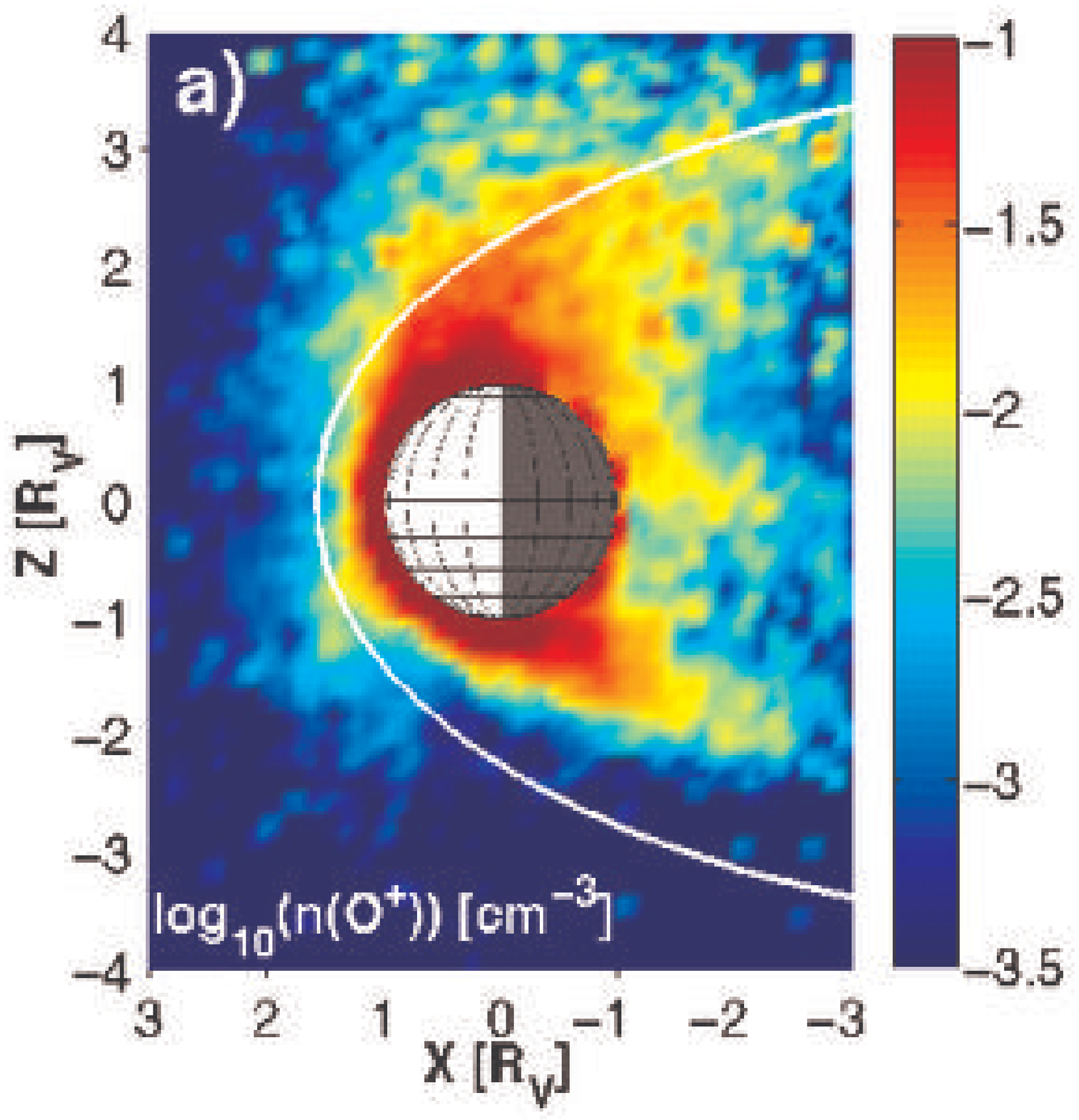}
  \includegraphics[width=0.24\textwidth]{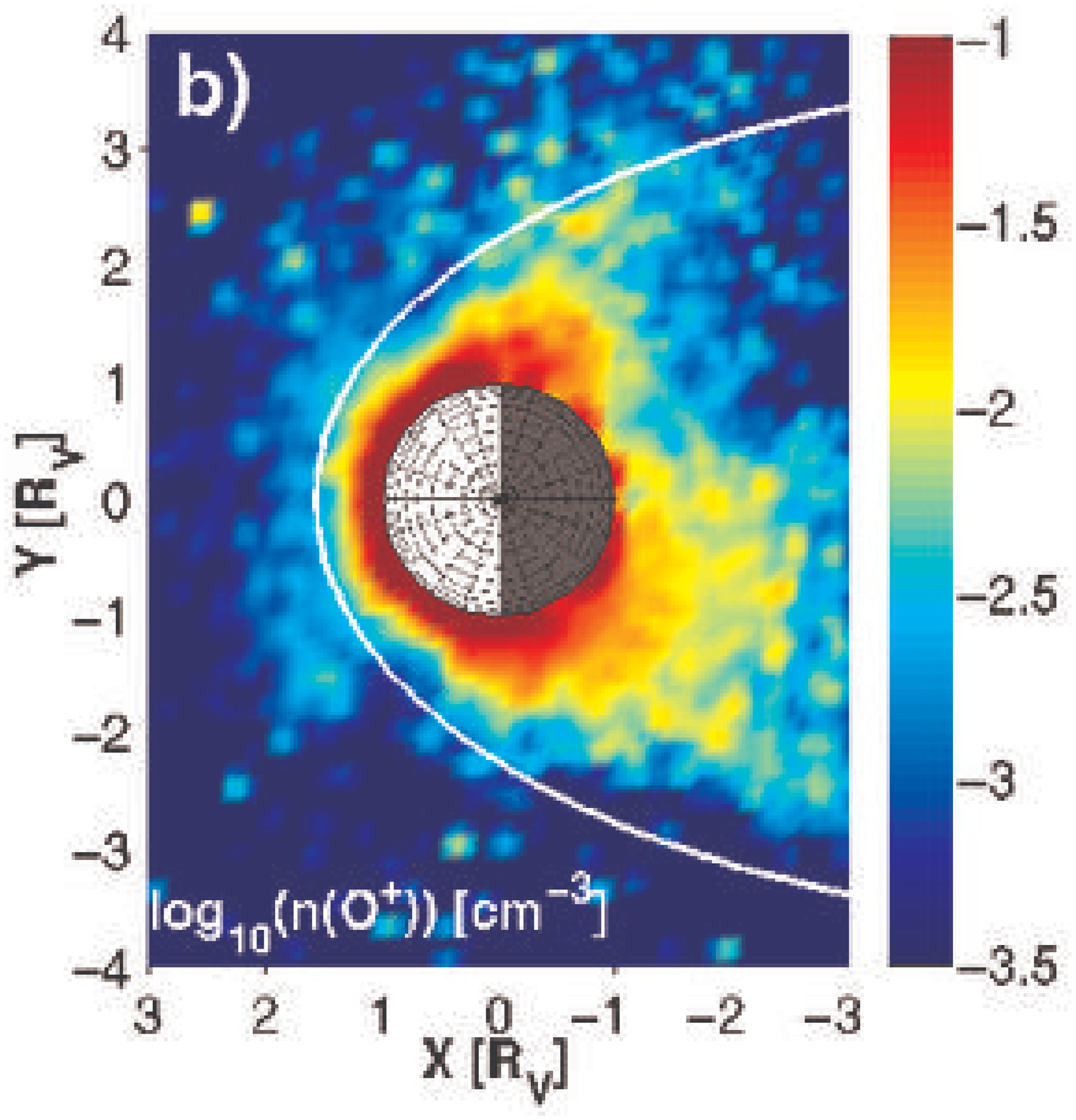}\\
  \includegraphics[width=0.24\textwidth]{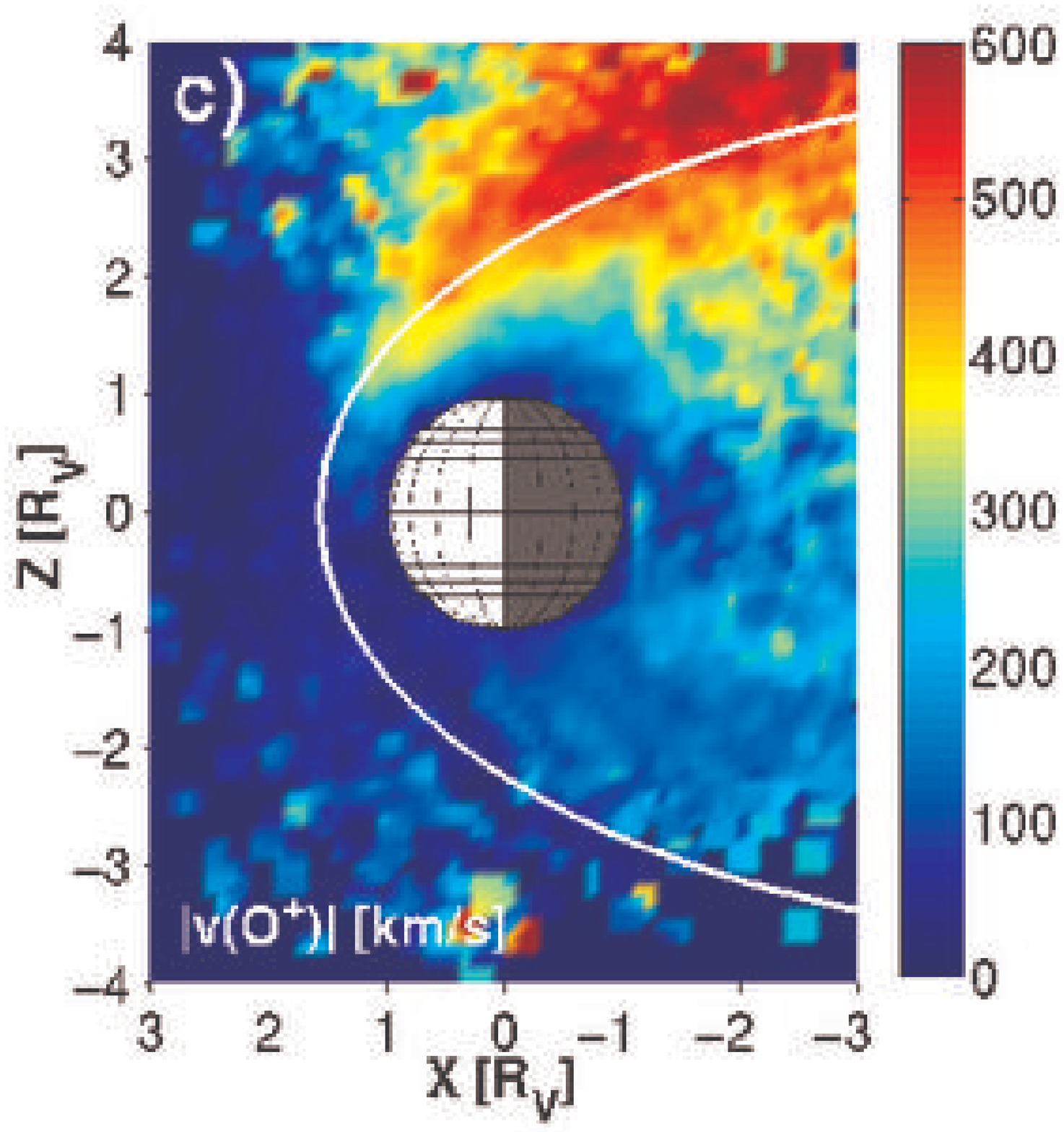}
  \includegraphics[width=0.24\textwidth]{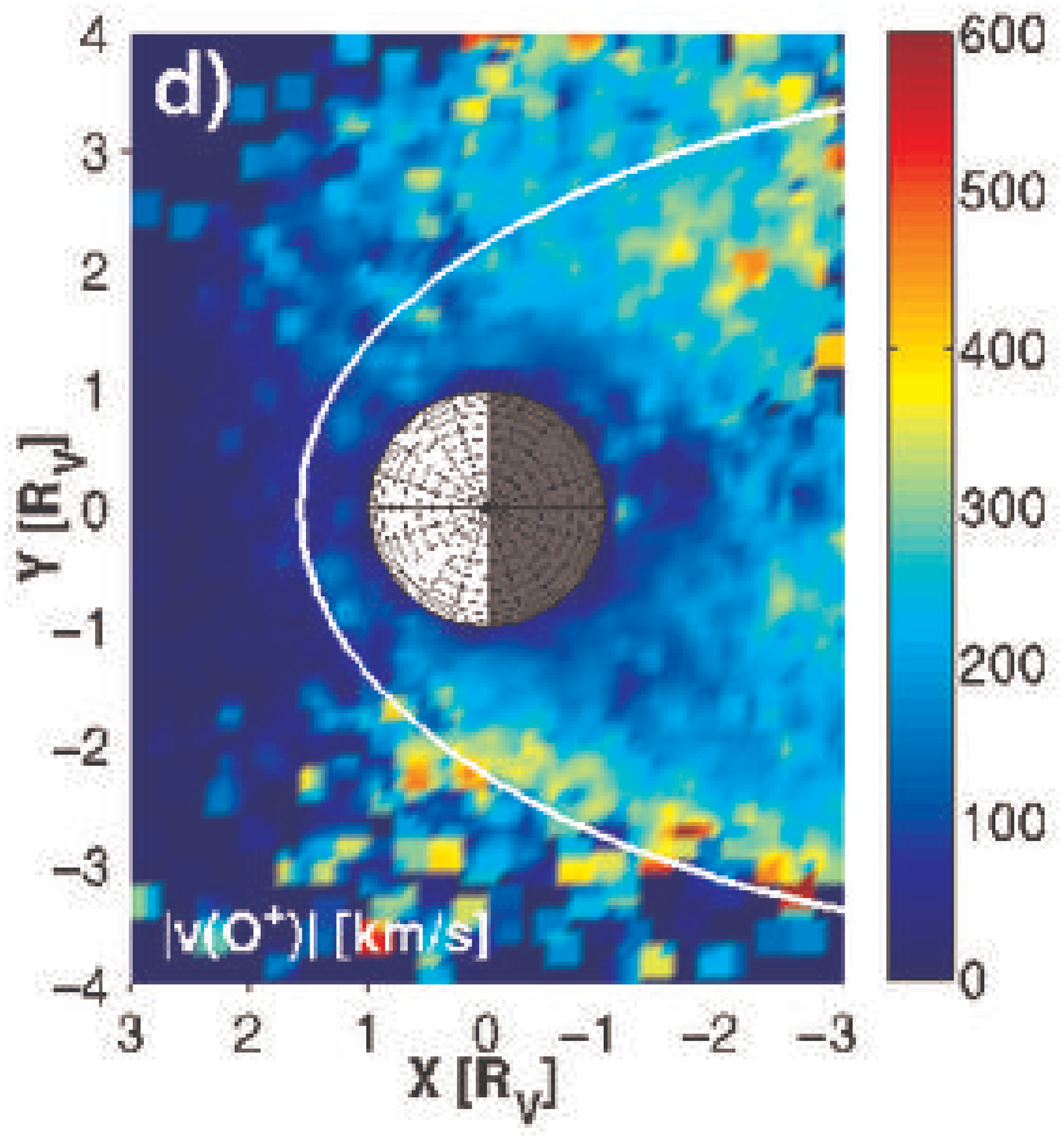}
  \caption{Properties of Venusian O$^+$ ions: Particle density [$\log_{10}$([cm$^{-3}$])] (a) in $XZ$ plane and (b) in $XY$ plane and the bulk velocity [km s$^{-1}$] (c) in $XZ$ plane and (d) in $XY$ plane.} \label{pic:O_xz_xy}
 \end{center}
\end{figure*}

Figure \ref{pic:O_xz_xy} presents plasma parameters similar to those in Fig. \ref{pic:H_xz_xy} but now for the oxygen ions. In this case a very clear +Esw/-Esw asymmetry is found. On the -Esw hemisphere a much sharper density gradient exists between the bow shock and the planet than on the +Esw hemisphere. Also, the bulk velocity of O$^+$ ions near the bow shock and in the solar wind is much higher on the +Esw hemisphere than on the opposite hemisphere. In addition, the density of the oxygen ions near Venus is higher on the -Esw hemisphere than on the +Esw hemisphere. These asymmetries can be understood on the basis of the direction of the convective electric field and the large gyroradius of O$^+$ ions compared to the size of the planet: the convective electric field accelerates newly born O$^+$ ions away (against) from the planet on the +Esw(-Esw) hemisphere in the $XZ$ plane. A slight BS$_{perp}$/BS$_{par}$ asymmetry arises in $n$(O$^+$) the O$^+$ density being slightly higher on the BS$_{perp}$ hemisphere than on the BS$_{par}$ hemisphere.

\begin{figure*}
 \begin{center}
  \includegraphics[width=0.24\textwidth]{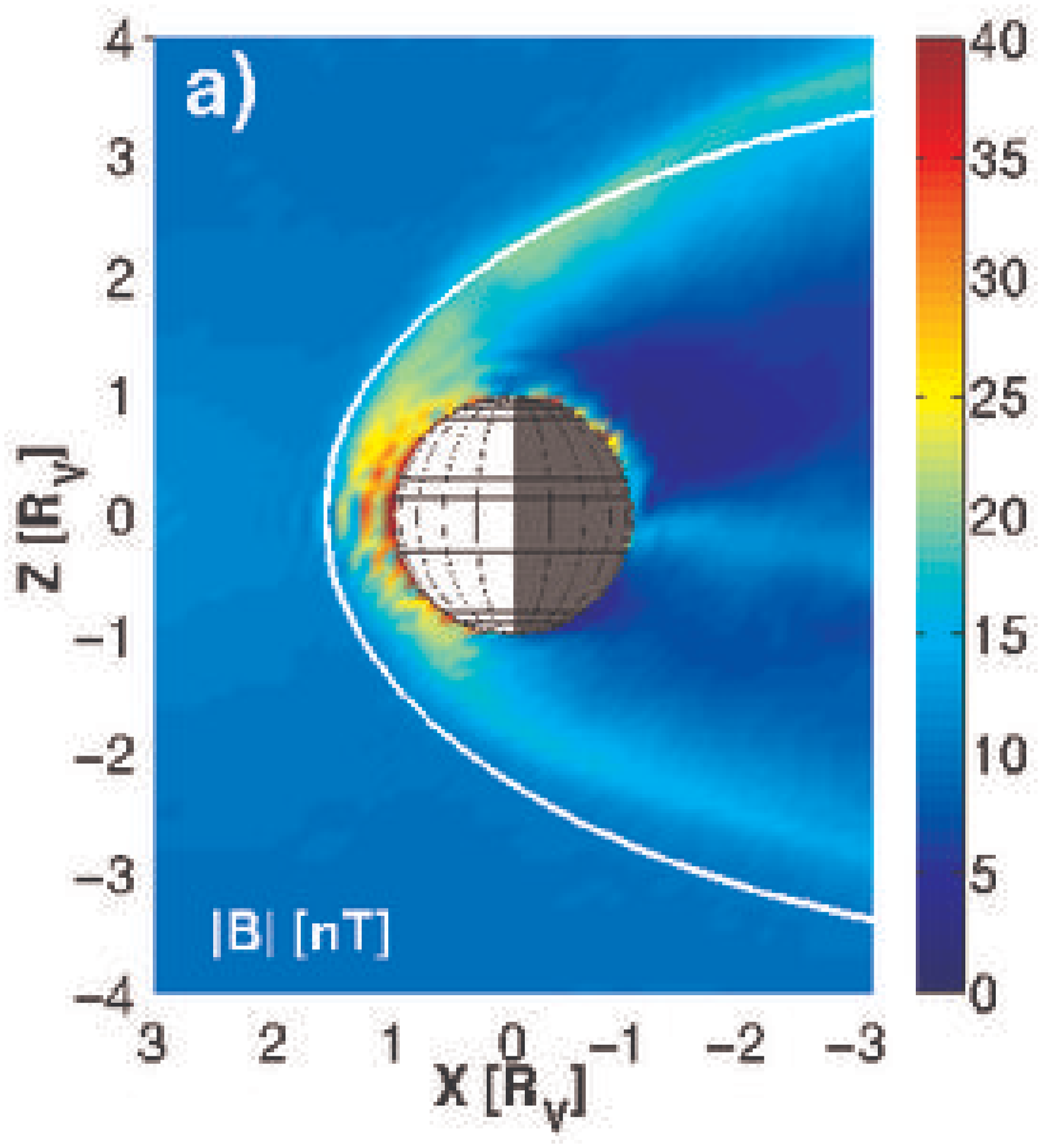}
  \includegraphics[width=0.24\textwidth]{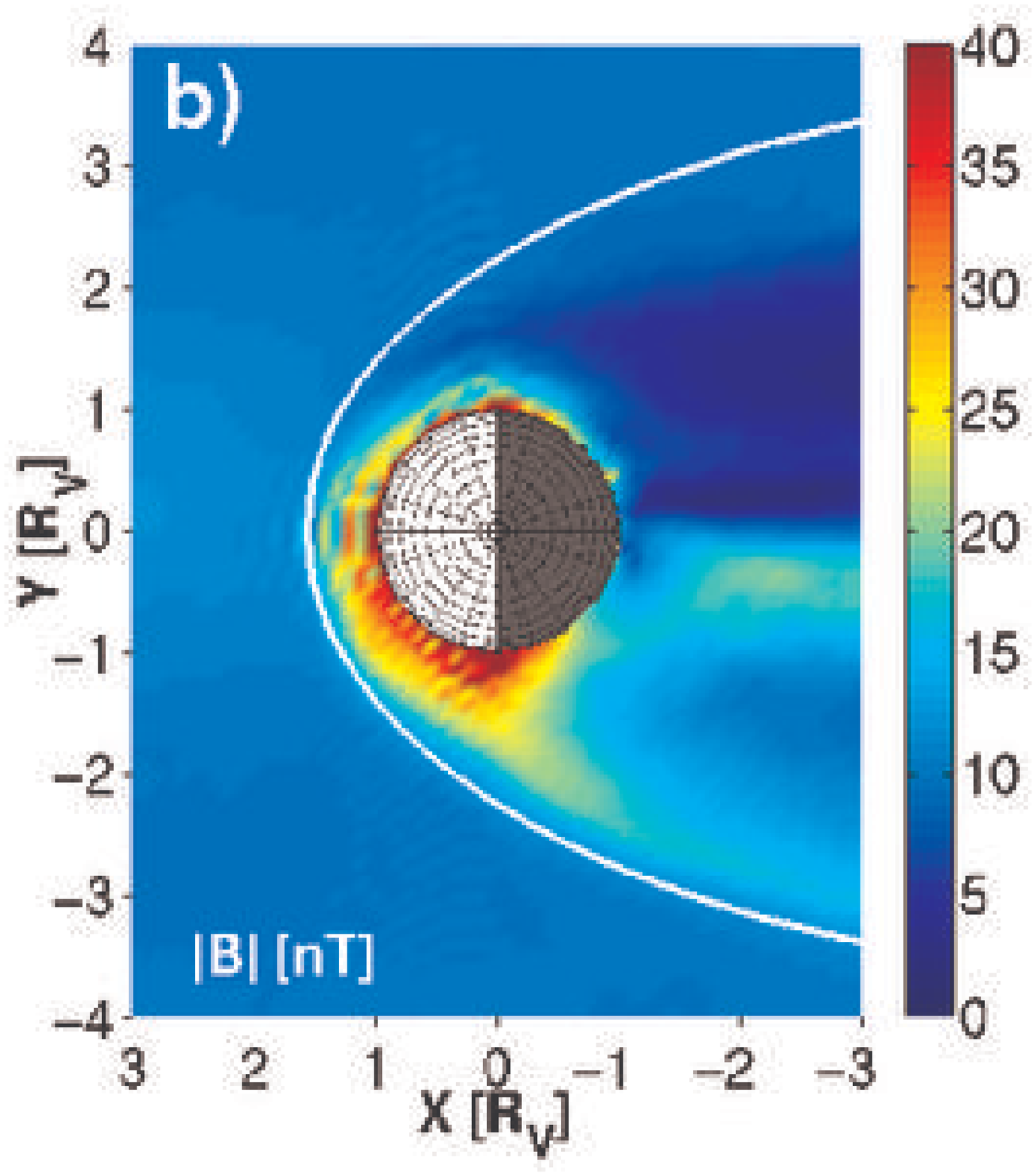}\\
  \includegraphics[width=0.24\textwidth]{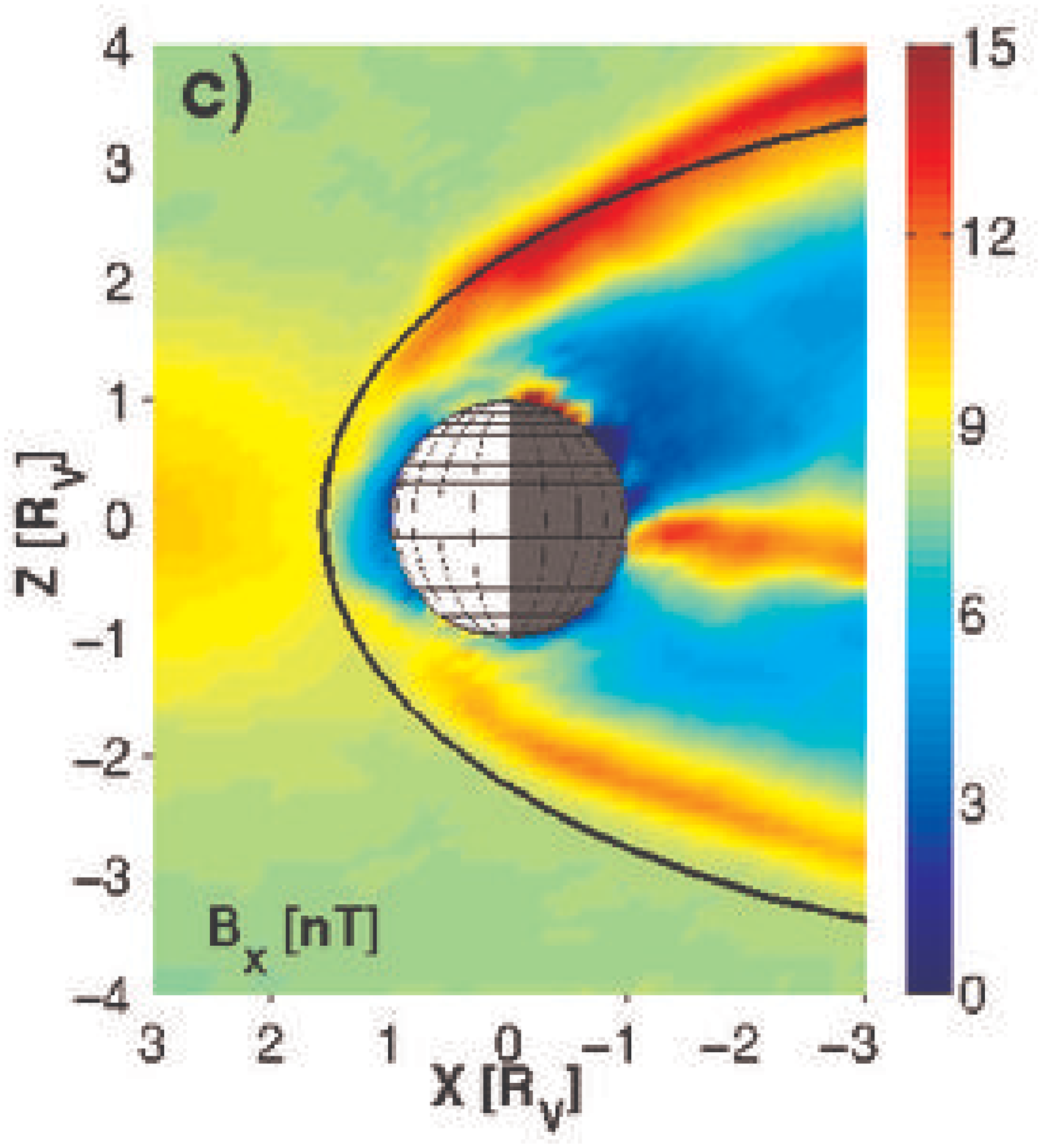}
  \includegraphics[width=0.24\textwidth]{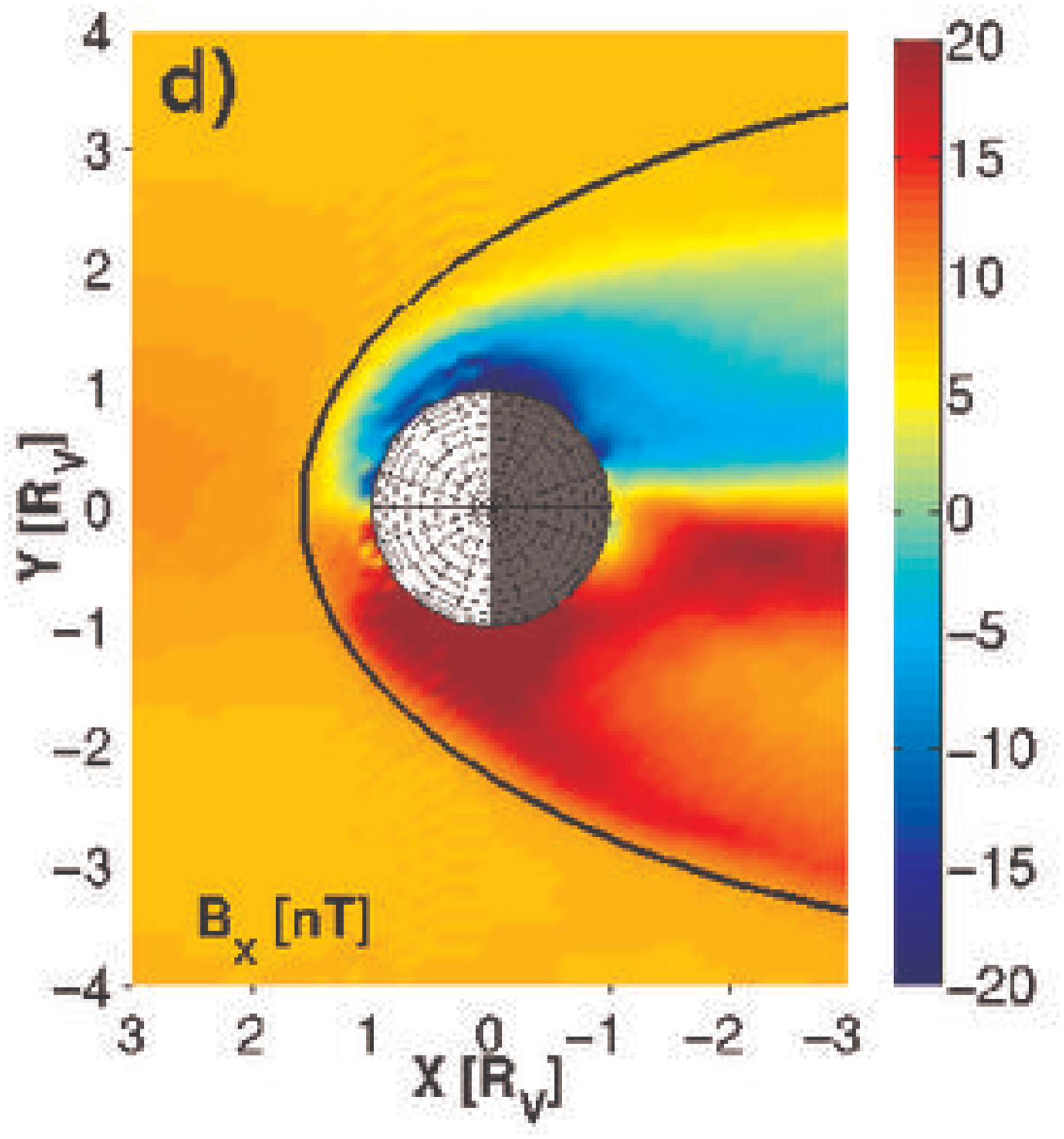}
  \caption{Magnetic field near Venus: The total magnetic field [nT] (a) in $XZ$ plane and (b) in $XY$ plane and the $x$-component [nT] (c) in $XZ$ plane and (d) in $XY$ plane.} \label{pic:B_xz_xy}
 \end{center}
\end{figure*}

Figure \ref{pic:B_xz_xy} shows how the plasma parameters are associated with the properties of the magnetic field. The bow shock can be identified clearly from an enhanced total magnetic field and enhanced $B_x$ component near the white line in Fig. \ref{pic:B_xz_xy}a and the black line in 6c, respectively. Increase of $B_x$ in the middle of the tail is associated with the magnetic tail lobe on the BS$_{perp}$ hemisphere, as will be seen later.

In the $XY$ plane a clear BS$_{perp}$/BS$_{par}$ asymmetry can be found. The total magnetic field is much larger on the BS$_{perp}$ hemisphere ($y < 0$) than on the opposite hemisphere. In addition, the magnitude of $B_x$ on the magnetic lobe on the BS$_{perp}$ hemisphere is much larger than $|B_x|$ on the opposite hemisphere. The asymmetry is associated with the IMF $x$-component: $x$-component of the $\widetilde{\vec{B}}_{laminar}$ in Eq. \ref{eq:Btotal} is positive and thus it decreases $|B_x|$ in the BS$_{par}$ hemisphere where $B_x$ is negative in the magnetic lobe. On the BS$_{perp}$ hemisphere $B_x$ is positive in the magnetic lobe and on that hemisphere $\widetilde{\vec{B}}_{laminar}$ increases $|B_x|$.

\begin{figure*}
 \begin{center}
  \includegraphics[width=0.24\textwidth]{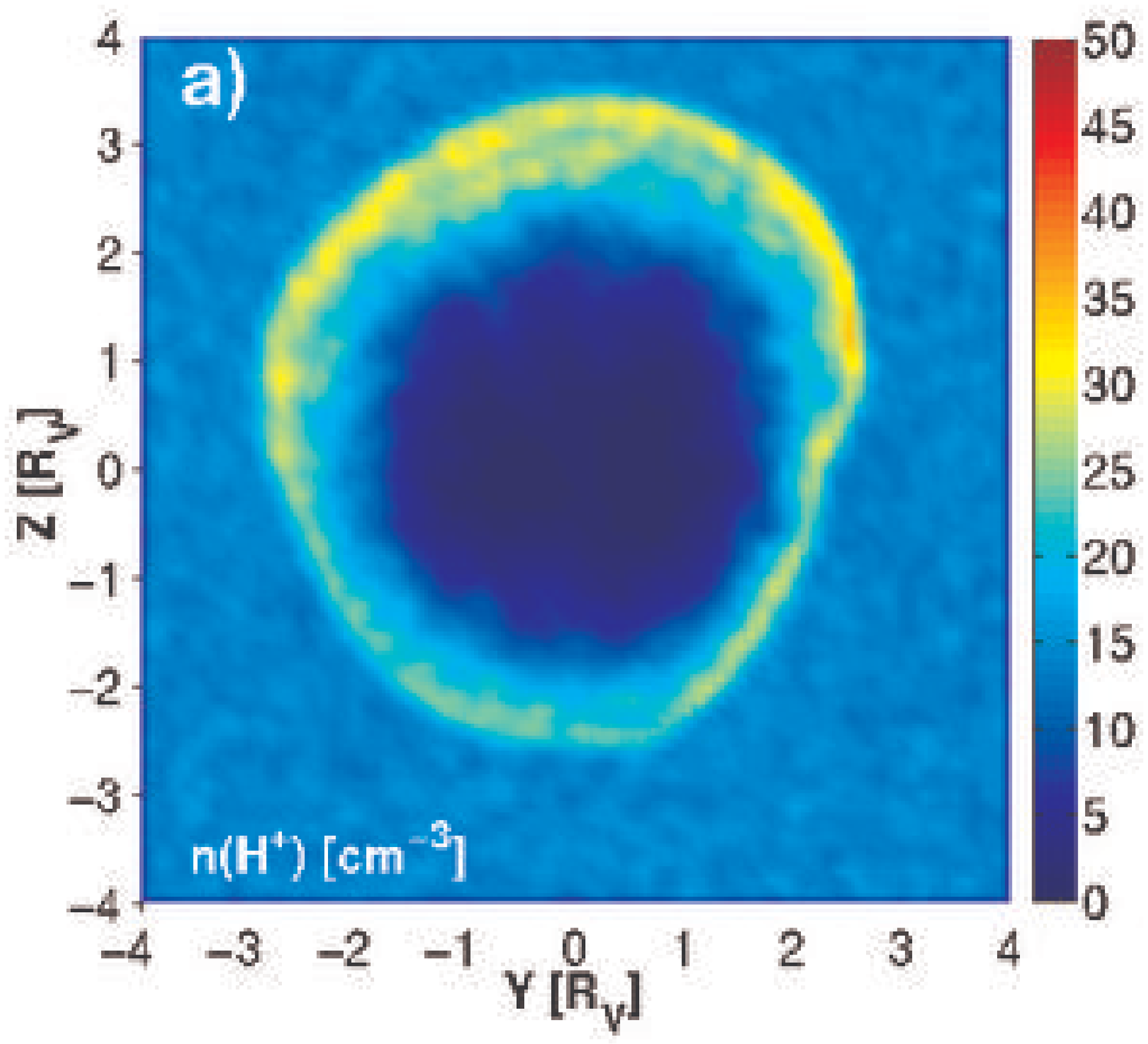}
  \includegraphics[width=0.24\textwidth]{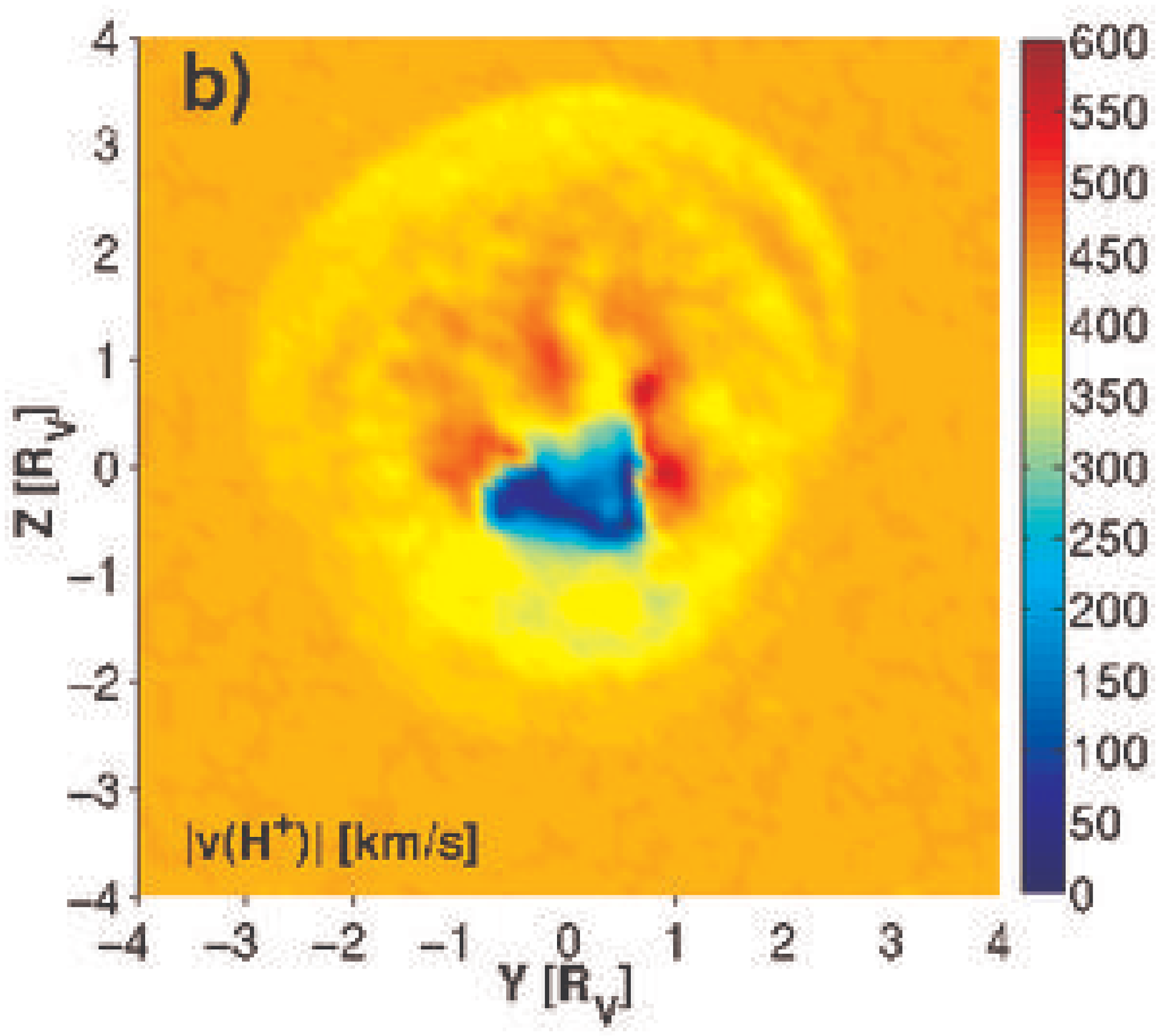}\\
  \includegraphics[width=0.24\textwidth]{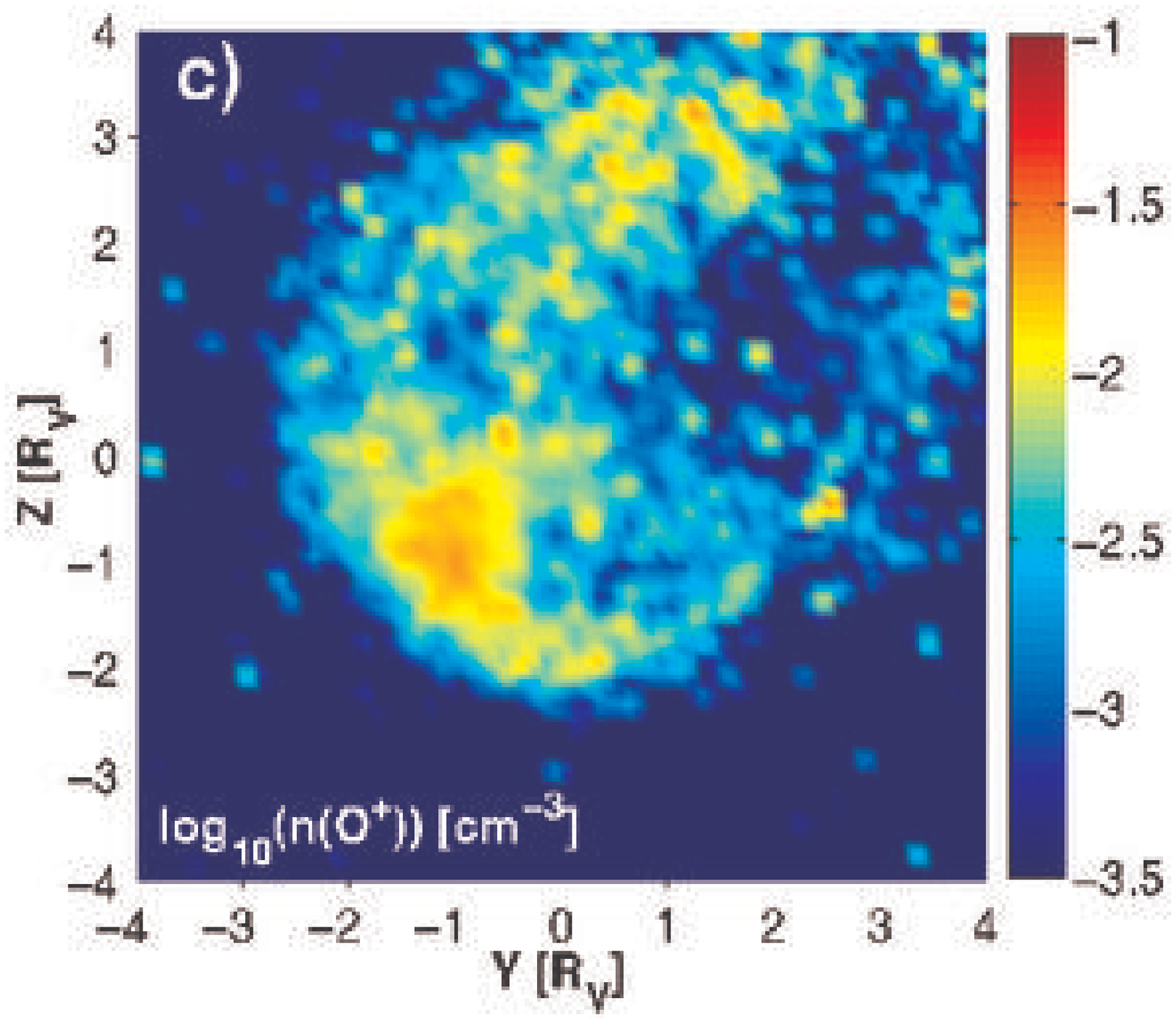}
  \includegraphics[width=0.24\textwidth]{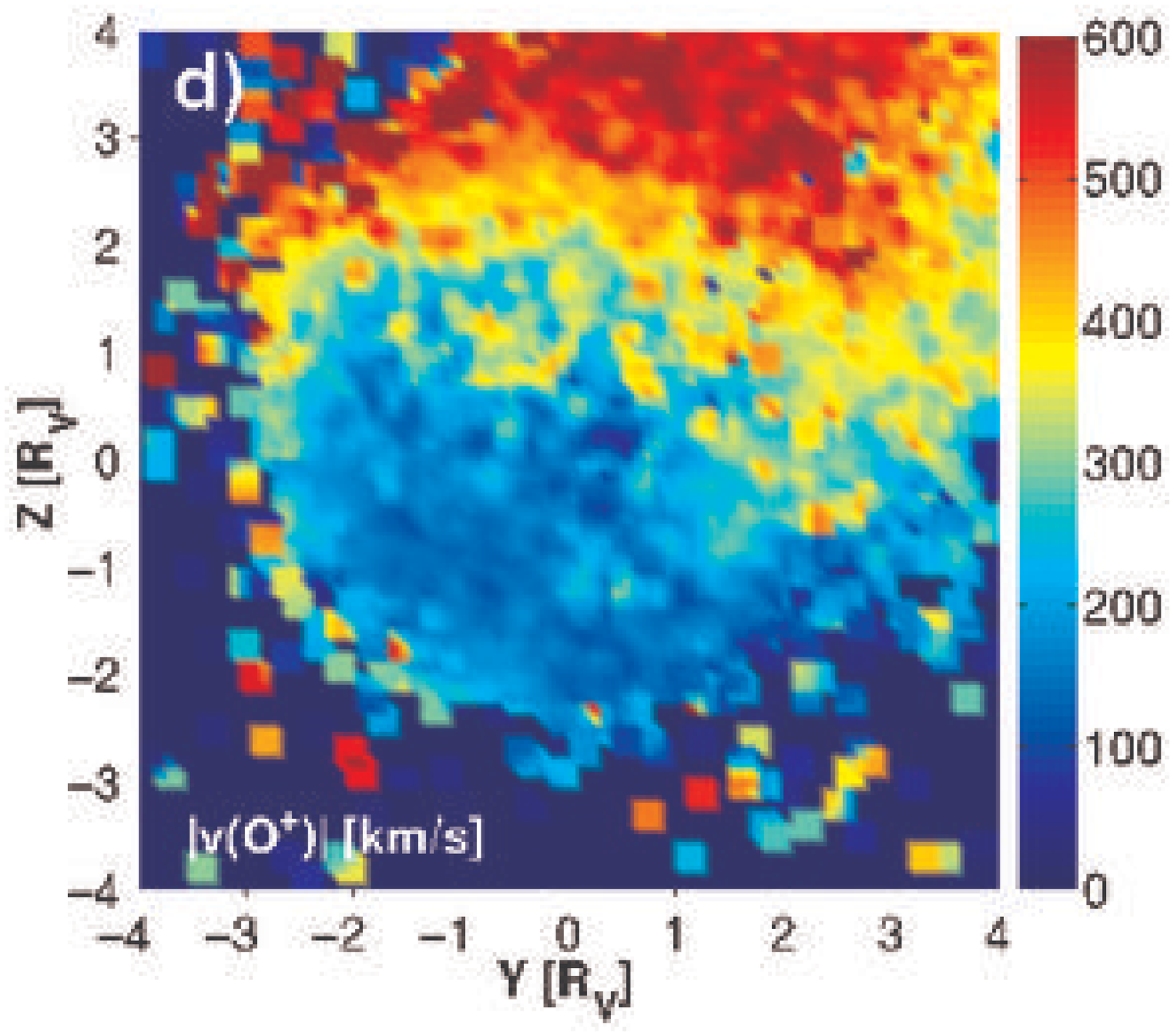}\\
  \includegraphics[width=0.24\textwidth]{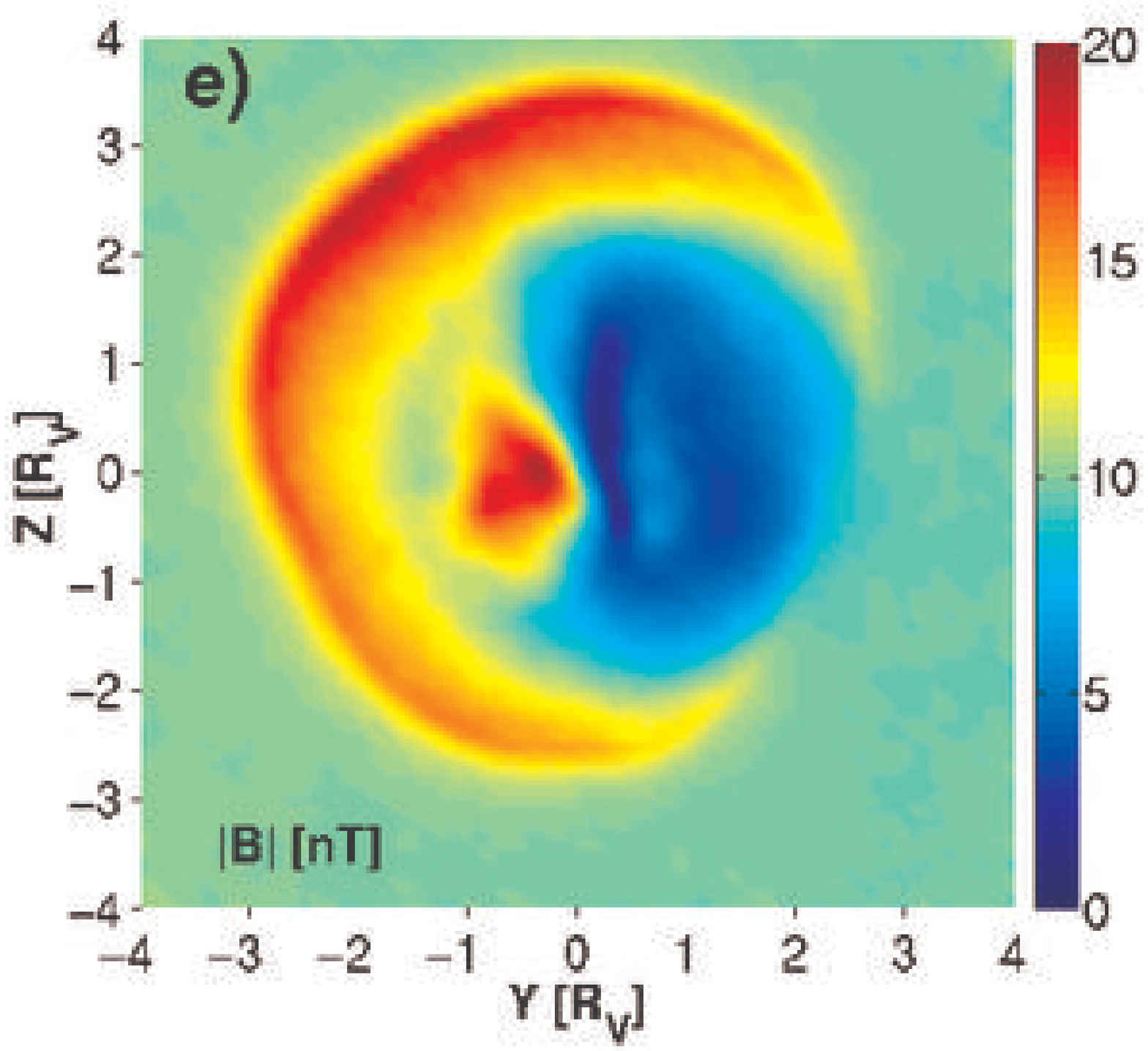}
  \includegraphics[width=0.24\textwidth]{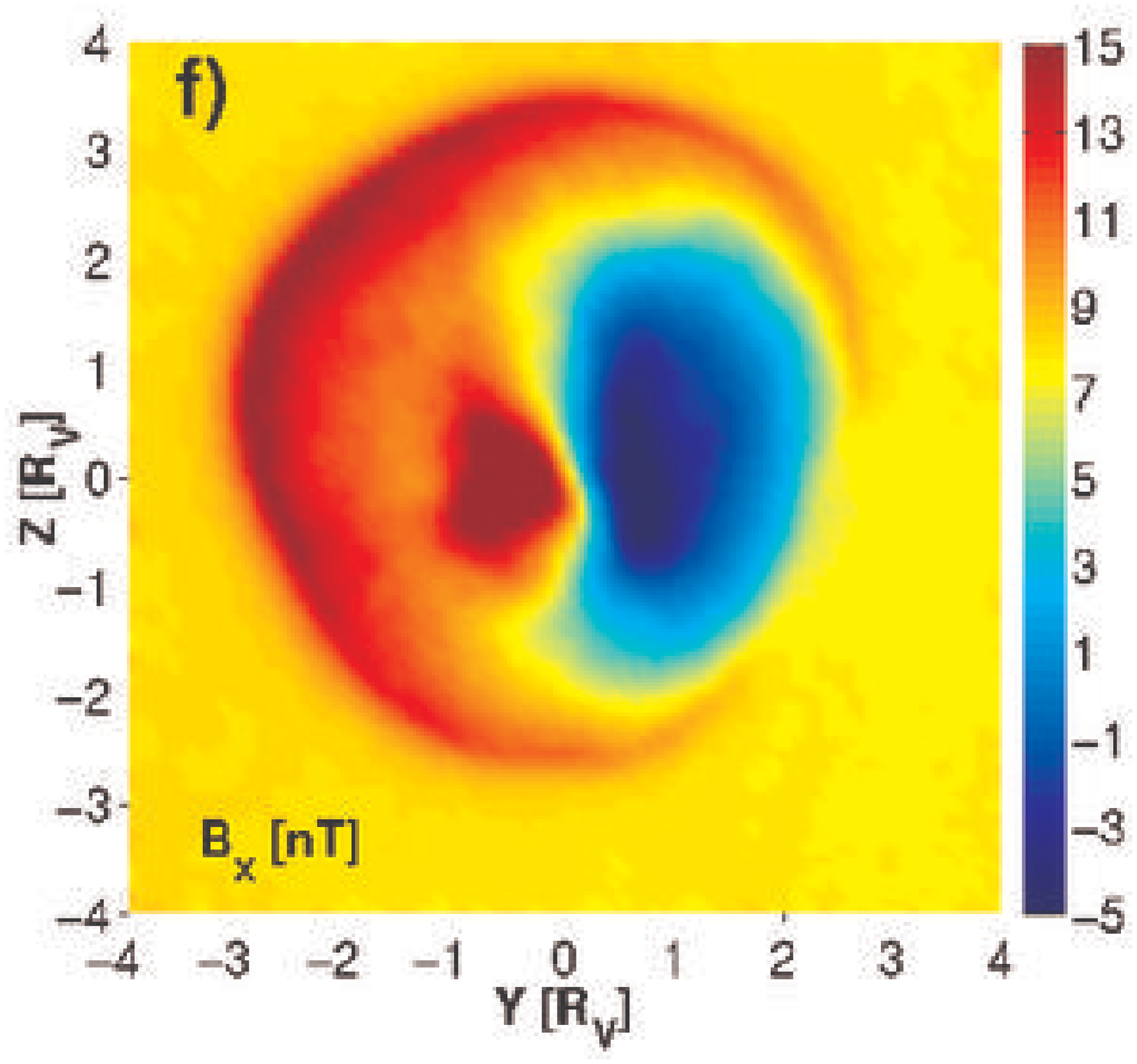}
  \caption{Plasma and the magnetic field at $x = -2 R_V$ plane: (a) the density [$n$(H$^+$) in cm$^{-3}$] and (b) the velocity [$v$(H$^+$) in km/s$^{-1}$] of protons, (c) the density [$n$(O$^+$) in $\log_{10}$([cm$^{-3}$]) scale] and (d) the velocity [$v$(O$^+$) in km s$^{-1}$] of the escaping planetary O$^+$ ions, (e) the total magnetic field [$|B|$ in nT], and (f) the $x$-component of the magnetic field [$B_x$ in nT].} \label{pic:yz}
 \end{center}
\end{figure*}

Plasma and magnetic field in the Venusian tail is studied in more detail in Fig. \ref{pic:yz}. Note that the density of O$^+$ ions is in its maximum downstream of the bow shock in the same region where the proton density is low. The +Esw/-Esw asymmetry can be seen in all plasma parameters: the bow shock is farther away from the planet in $z > 0$ than in $z < 0$ (Fig. \ref{pic:yz}a), the proton velocity is higher in $z > 0$ than in the opposite hemisphere (Fig. \ref{pic:yz}b), the maximum $n$(O$^+$) is sited in the $z < 0$ hemisphere and O$^+$ ions can be found in the solar wind in the $z > 0$ hemisphere (Fig. \ref{pic:yz}c), the O$^+$ ions obtain higher velocities in the +Esw hemisphere ($z > 0$) than in the opposite hemisphere (Fig. \ref{pic:yz}d). The magnetic field is also asymmetric with respect to the direction of $\vec{E}_\textrm{SW}$.

A somewhat unexpected feature in Figures \ref{pic:yz}e and \ref{pic:yz}f is the very clear asymmetry between the magnetic tail lobes. In the analyzed IMF direction case the electric currents in the tail results in the magnetic tail lobes having $B_x > 0$ at $y < 0$ and $B_x < 0$ at $y > 0$. As already noted before, the IMF $x$-component, which is modelled by $\widetilde{\vec{B}}_{laminar}$ in this study, decreases the total magnetic field at the $y > 0$ hemisphere and increases the field at the opposite hemisphere. The magnetic tail lobe is therefore much stronger at $y < 0$ than at $y > 0$.

A similar asymmetry can be found in the strength of the magnetic field between the $y < 0$ and $y > 0$ hemispheres. The magnetic field is clearly enhanced at the bow shock and in the magnetosheath in the BS$_{perp}$ hemisphere ($y < 0$). At the BS$_{par}$ hemisphere ($y > 0$), instead, the enhancement of the field is much weaker. In fact, practically no enhancement of the magnetic field can be seen in the BS$_{par}$ hemisphere at the $z = 0$ plane.

\begin{figure}
 \begin{center}
  \includegraphics[width=0.49\textwidth]{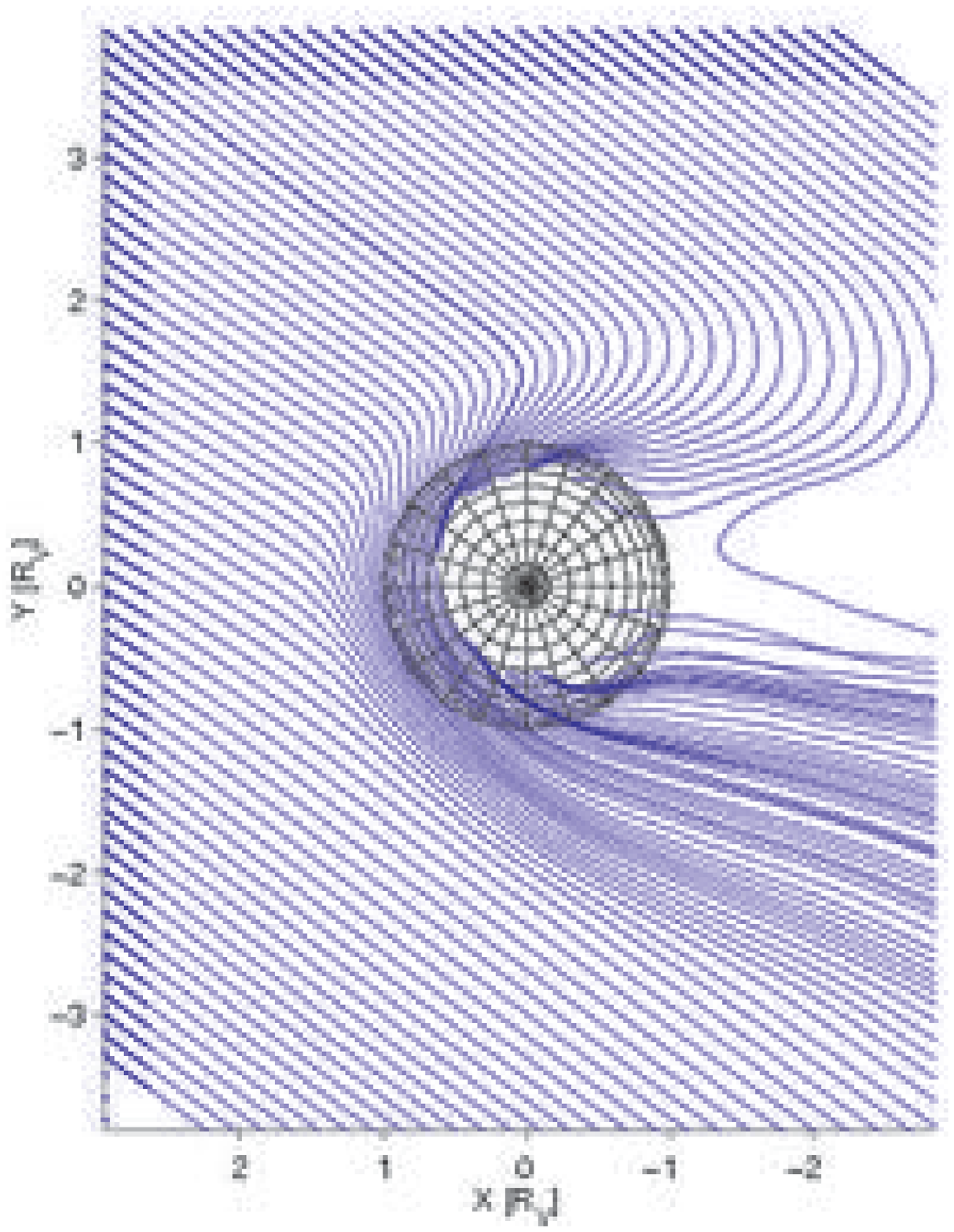}
  \caption{Draped magnetic field lines at Venus. Note the asymmetry between the quasi-perpendicular bow shock hemisphere, BS$_{perp}$ ($y < 0$), and the quasi-parallel bow shock hemisphere, BS$_{par}$ ($y > 0$).} \label{pic:draping}
 \end{center}
\end{figure}

Figure \ref{pic:draping} illustrates how the morphology of the magnetic field is associated with the BS$_{perp}$/BS$_{par}$ asymmetry. On the BS$_{perp}$ hemisphere magnetic field piles up against the magnetic tail lobe, resulting in an enhanced magnetic field. On the BS$_{perp}$ hemisphere the sign of the $B_x$ at the magnetic tail lobe is the same as the sign of the $B_x$ in the solar wind. On the opposite BS$_{par}$ hemisphere the sign of the $B_x$ in the magnetic tail lobe is opposite to the sign of the $B_x$ in the solar wind and the direction of the field line therefore must change.

\section{Discussion}
This paper presents the first step to study the Venusian plasma environment with a new 3-D QNH model. To the authors' knowledge it is the first published QNH model study which uses a realistic Parker spiral IMF where the IMF $x$-component is larger in magnitude than the magnetic field components perpendicular to the flow.

The study suggests that the model can reproduce the basic observed plasma and magnetic field regions and boundaries near Venus, implying a future potential of the developed numerical approach. However, the study leaves many open issues for future studies. The grid resolution is not fine enough to resolve the position and the shape of the bow shock and the inner structure of the magnetic barrier. Especially, the grid resolution makes it impossible to include a self-consistent ionosphere. Furthermore, at the present stage of the model the total O$^+$ emission rate from the obstacle boundary is a freely chosen parameter. In the future, its value can be adjusted by comparing the simulated O$^+$ densities with the observed densities. Similar approach has been used earlier to estimate the total ion escape rate at Mars and a similar type of estimation can be made when direct O$^+$ measurements from ASPERA-4/Venus Express mission becomes available in 2006. In the future one of the biggest challenges in modelling is to implement a more realistic ionosphere to the 3-D QNH model.

The second limitation of the presented study is that is presents the solution only for one set of upstream parameters. In practice, Venus is under the influence of varying plasma and magnetic field conditions \citep[see, for example,][]{LuhmannEtAl_PVO_1993}. How the model responds to the upstream parameters, which also have time dependence, will be a topic of future studies.

It is worth recalling that in the presented run the resistivity is constant outside Venus. The used resistivity was not 'optimized' to its minimum value and the value is artificial in the sense that its value can be smaller in reality. The specific resistivity value was adopted because it has been found to be large enough to be used for various upstream parameter runs (not shown in this paper) to reduce fluctuations formed near the obstacle boundary (Venus). The role of the non-uniform, for example, spherically symmetric resistivity model, will be studied in the future when a smaller grid size near Venus than used in this paper will be used.

It is also worth noting that there is a large magnetic field data set available from Pioneer Venus Orbiter (PVO) over ten year measurements. The data set makes it possible to make a detailed quantitative comparison between the developed QNH model and the observations. In fact, preliminary comparisons between the model and the magnetic field measurements made by PVO when the spacecraft clearly crossed both magnetic tail lobes \citep[see][]{LuhmannEtAl_Induced_1991} have shown that the developed Venus model is capable of reproducing many of the observed features (figures not shown). In the model the high velocity O$^+$ ions were also found on the +Esw hemisphere (c.f. Fig. \ref{pic:O_xz_xy}c), that is, on the same hemisphere where PVO observed fast moving escaping O$^+$ ions, for example, in the far tail \citep[see, for example,][]{SlavinEtAl_PVO_1989}. A detailed comparison between PVO magnetic field and ion observations and the model will be a topic of future studies.

It is interesting to note that the BS$_{perp}$/BS$_{par}$ asymmetry of the bow shock on the $XY$ plane (Fig. \ref{pic:H_xz_xy}b and \ref{pic:H_xz_xy}d) is in agreement with observations \citep[see][]{ZhangEtAl_BS_1991}. One may anticipate that a detailed comparison between the data and the model will provide a new insight about the applicability of the global 3-D QNH model to reproduce plasma parameters at various upstream conditions. The developed QNH model can be used to separate effects of different kinds. In some sense the model can also be regarded to provide a possibility to "filter" away from the measurements such effects that result from the very basic laws of nature (Lorentz force, Faraday's and Amp\`{e}re's law and the conservation of the electron's momentum) under the used boundary conditions from effects that possibly result from unknown factors not implemented in the model. For example, the role of the electron impact ionization and the role of the ion-neutral collisions remains to be studied, especially, which kind of global effects the omitted process may cause.

Finally, possible artefacts result from the adopted magnetic field model (Equations \ref{eq:laminarB}) have to be studied by comparing runs made for different magnetic field configurations.

\section{Summary}
A three dimensional self-consistent numerical model is developed to study the Venus-solar wind interaction. An analysis of the runs made for a Parker spiral IMF direction suggests that several asymmetries take place at Venus. The observed +Esw/-Esw asymmetry results from ion finite gyroradius effects and from the escaping O$^+$ ions while the IMF $x$-components caused BS$_{perp}$/BS$_{par}$ asymmetry. The density of the escaping planetary O$^+$ ions in the magnetosheath and in the solar wind is found to be higher in the hemisphere where the convective electric field points in the undisturbed solar wind than in the opposite hemisphere. The analysis implies that the properties of plasma and magnetic field at given points near Venus can be anticipated to vary noticeably in concert with varying upstream parameters.

\end{document}